\documentclass[reqno,centertags, letterpaper,12pt]{amsart}  
\usepackage{amsmath,amsthm,amscd,amssymb}
\usepackage{latexsym}
\numberwithin{equation}{section}

\newtheorem{theorem}{Theorem}[section]

\newtheorem{lemma}[theorem]{Lemma}

\newtheorem{proposition}[theorem]{Proposition}

\theoremstyle{definition}
\newtheorem*{example}{Example}
\newtheorem*{ttt}{Theorem}

\newcommand{\bi}{\bibitem}
\newcommand{\lb}{\label}

\newcommand{\bbR}{{\mathbb{R}}}
\newcommand{\bbN}{{\mathbb{N}}}
\newcommand{\bbZ}{{\mathbb{Z}}}
\newcommand{\bbQ}{{\mathbb{Q}}}
\newcommand{\bbE}{{\mathbb{E}}}
\newcommand{\calK}{{\mathcal{K}}}
\newcommand{\calX}{{\mathcal{X}}}
\newcommand{\calE}{{\mathcal{E}}}
\newcommand{\calO}{{\mathcal{O}}}

\newcommand{\ga}{{\gamma}}
\newcommand{\del}{{\delta}}
\newcommand{\park}{{\frac{\partial}{\partial k}}}
\newcommand{\tht}{{\theta}}
\newcommand{\thh}{{\overline{\theta}}}
\newcommand{\eps}{{\varepsilon}}
\newcommand{\sgn}{\text{\rm{sgn}}}
\newcommand{\dist}{\text{\rm dist}}
\newcommand{\res}{|}
\newcommand{\lims}{\operatorname*{\varlimsup}}
\newcommand{\limi}{\operatorname*{\varliminf}}

\begin{document}

\title[Sparse Potentials With Fractional Dimension]{Sparse Potentials With Fractional Hausdorff Dimension}
\author[Andrej Zlato\v s]{Andrej Zlato\v s}

\address{Mathematics 253-37 \\
California Institute of Technology \\
Pasadena, CA 91125 }

\email{andrej@caltech.edu}


\date{28 October, 2002}

\begin{abstract}
We construct non-random bounded discrete half-line Schr\" odinger operators which have purely singular continuous spectral measures with fractional Hausdorff dimension (in some interval of energies). To do this we use suitable sparse potentials. Our results also apply to whole line operators, as well as to certain random operators. In the latter case we prove and compute an exact dimension of the spectral measures.
\end{abstract}

\maketitle


\section{Introduction} \lb{S1}

In the present paper, we consider discrete half-line Schr\" odinger operators $H_{\phi}$ on $\ell^2(\bbZ^+)=\ell^2(\{1,2,\dots\})$,  given by
\begin{equation} \lb{1.1}
(H_{\phi}u)(x) \equiv u(x+1)+u(x-1)+V(x)u(x)
\end{equation}
for $x\in\bbZ^+$, with the potential $V$ and boundary condition $\phi\in(-\tfrac \pi 2,\tfrac \pi 2]$
\begin{equation} \lb{1.2}
u(0)\cos(\phi)+u(1)\sin(\phi)=0.
\end{equation}
Here \eqref{1.2} defines $u(0)$, which then enters in \eqref{1.1} for $x=1$. Hence $H_0$ is the {\it Dirichlet operator\/} with $u(0)=0$, and $H_{\phi}=H_0-\tan(\phi)\delta_1$ where $\delta_1$ is the delta function at $x=1$. $H_{\pi/2}$ is the {\it Neumann operator\/} with $u(1)=0$. All these are rank one perturbations of $H_0$.

A function $u$ on $\bbZ^+\cup\{0\}$ is a {\it generalized eigenfunction\/} of the above operators for energy $E$ and boundary condition $\phi$ if 
\begin{equation} \lb{1.3}
u(x+1)+u(x-1)+V(x)u(x)=Eu(x)
\end{equation}
for $x\in\bbZ^+$ and \eqref{1.2} holds. Since such $u$ is uniquely given by its values at $x=0,1$, the space of generalized eigenfunctions for any energy is 2-dimensional. The $2\times 2$ unimodular matrix $T_E(x,y)$ which takes  $\binom{u(y+1)}{u(y)}$ to $\binom{u(x+1)}{u(x)}$ whenever $u$ is a generalized eigenfunction for energy $E$, is called the {\it transfer matrix} for $E$. It is immediate that
\[
T_E(x,y)=\prod_{j=y+1}^x T_E(j,j-1)=\prod_{j=y+1}^x \begin{pmatrix} E-V(j)&-1\\1&0\end{pmatrix}.
\]
We denote $T_E(x)\equiv T_E(x,0)$.

We let $\mu_{\phi}$ be the spectral measures of the above operators. The aim of this paper is to construct a (non-random) bounded potential $V$ such that these measures are purely singular continuous and have fractional (not 0 or 1) Hausdorff dimension in some interval of energies. We consider the sparse potential with equal barriers $V_{v,\ga}$ given by \eqref{1.6} below, with $v\neq 0$ and $\ga\ge 2$. Here is our main result:

\begin{ttt} \lb{T1.0}
Let $H_{\phi}$ be the discrete Schr\" odinger operator on $\bbZ^+$ with potential $V_{v,\ga}$ given by \eqref{1.6}, and boundary condition $\phi$. Let $\mu_{\phi}$ be its spectral measure. For any closed interval of energies $J\subset(-2,2)$ there is $v_0>0$ and $\ga_0\in\bbN$ such that if $0<|v|<v_0$ and $\ga\ge\ga_0v^{-2}$ is an integer, then for any $\phi$, the measure $\mu_{\phi}$ has fractional Hausdorff dimension in $J$.
\end{ttt}

This is proved in Section \ref{S5} as {\it Theorem \ref{T5.1}}. From the rest of our results we would like to single out {\it Theorems \ref{T1.4}, \ref{T6.3}\/} (random case) and {\it \ref{T7.1}}.

Our motivation is a paper by Jitomirskaya-Last \cite{JL}, which relates the power growth/decay of eigenfunctions and the Hausdorff dimension of spectral measures. We will apply ideas from \cite{KLS} and use sparse potentials, which allow us to control this growth. We mention that \cite{JL} also provides an example of potentials with the above properties, but these are unbounded (and hence so are the operators). 

First we recall some basic facts about dimension of sets and measures. If $S\subseteq\bbR$ and $\alpha\in[0,1]$, then the $\alpha$-dimensional Hausdorff measure of $S$ is
\[
h^\alpha(S)\equiv 
\lim_{\delta\to 0} \,\bigg[ \inf_{\delta\text{-covers}} \, \sum_{n=1}^\infty |I_n|^\alpha \bigg] .
\]
Here a $\delta$-cover is a covering of $S$ by a countable set of intervals $I_n$ of lengths at most $\delta$. Notice that $h^0$ is the counting measure and $h^1$ the Lebesgue measure. For any $S$ there is a number $\alpha_S\in[0,1]$ such that $h^\alpha(S)=0$ if $\alpha>\alpha_S$, and $h^\alpha(S)=\infty$ if $\alpha<\alpha_S$. This $\alpha_S$ is the {\it dimension\/} of $S$. 

If $\mu$ is a measure on $\bbR$, we say that $\mu$ is {\it $\alpha$-continuous\/} if it is absolutely continuous with respect to $h^\alpha$, and $\mu$ is {\it $\alpha$-singular\/} if it is singular to $h^\alpha$. Hence $\alpha$-continuous measures do not give weight to sets $S$ with $h^\alpha(S)=0$ (e.g., to sets $S$ such that $\dim(S)<\alpha$), and $\alpha$-singular measures are supported on sets $S$ with $h^\alpha(S)=0$ (and so $\dim(S)\le \alpha$). We say that $\mu$ has {\it fractional Hausdorff dimension\/} in some interval $I$ if $\mu(I\cap\cdot)$ is $\alpha$-continuous and $(1-\alpha)$-singular for some $\alpha>0$. Finally, $\mu$ has {\it exact\/} ({\it local\/}) {\it dimension\/} in $I$ if for any $E\in I$ there is an $\alpha(E)$, and for any $\eps>0$ there is $\delta>0$ such that $\mu((E-\delta,E+\delta)\cap\cdot)$ is both $(\alpha(E)-\eps)$-continuous and $(\alpha(E)+\eps)$-singular. We do not prove an exact dimension for our measures $\mu_{\phi}$ (corresponding to \eqref{1.6}), but we do it for the random potential case which we consider in Section \ref{S6}.

In the present paper, we will sometimes say that $\alpha$-continuous measures have dimension at least $\alpha$ and that $\alpha$-singular measures have dimension at most $\alpha$.

We will mainly use two results from \cite{JL} ({\it Corollaries 4.4\/} and {\it 4.5\/}) which relate eigenfunction growth and spectral dimension. Here, however, these results will be restated in terms of the {\it EFGP transform\/} of eigenfunctions ({\it Propositions \ref{P1.2}\/} and {\it \ref{P1.3}\/} below), rather than in terms of the eigenfunctions themselves.

The EFGP transform $(R,\tht)$ of an eigenfunction $u$ with energy $E\in(-2,2)$ is a Pr\" ufer-type transform which makes the growth/decay of $u$ more transparent. It is defined as follows. We let $k\in(0,\pi)$ be such that $E=2\cos(k)$ and set
\begin{equation} \lb{1.4}
\begin{aligned}
u(x)-\cos(k)u(x-1) &=R(x)\cos(\tht(x)),
\\ \sin(k)u(x-1) &=R(x)\sin(\tht(x)).
\end{aligned}
\end{equation}
These equations define $R(x)>0$ and $\tht(x) \pmod {2\pi}$ uniquely, and we write $u\sim(R,\tht)$. 

If we set
\[
\thh(x)\equiv \tht(x)+k \qquad \text{and} \qquad v_k(x)\equiv -\frac{V(x)}{\sin(k)},
\]
then \eqref{1.3} becomes (see \cite{KLS})
\begin{equation} \lb{1.5}
\begin{gathered}
\cot\bigl(\tht(x+1)\bigr)=\cot\bigl(\thh(x)\bigr)+v_k(x),
\\ \frac{R(x+1)^2}{R(x)^2}
=1+v_k(x)\sin\bigl(2\thh(x)\bigr)+v_k(x)^2\sin^2\bigl(\thh(x)\bigr).
\end{gathered}
\end{equation}
Notice that \eqref{1.5} only determines $\tht(x+1)\pmod\pi$. There are two ways to deal with this. The first is to examine \eqref{1.4} more closely and conclude (as in \cite{KR}) that $\sgn\bigl(\sin(\tht(x+1))\bigr)=\sgn\bigl(\sin(\thh(x))\bigr)$, and if this is $0$, then $\sgn\bigl(\cos(\tht(x+1))\bigr)=\sgn\bigl(\cos(\thh(x))\bigr)$. This fact and \eqref{1.5} determine $\tht(x+1)\pmod {2\pi}$ uniquely, but we will not need this extra condition here. 

The second way is to realize that $\sin(2\tht)$, $\sin^2(\tht)$ and $\cot(\tht)$ all have period $\pi$, and so this ambiguity in $\tht$ does not affect the values of $R$, which are of main interest to us. 
Notice also, that once $u(0)$ and $u(1)$ are fixed and $k$ is varied, $u(x)$ is a polynomial in $E$ of degree $x$. Therefore $u(x)$, and by \eqref{1.4} also $R(x)$ and $\tht(x)$ (viewed as a function on the unit circle), are well-defined $C^\infty$ functions of $k$. Moreover, there is completely no ambiguity in $\park\tht(x)$, which will be of significant importance in our considerations. 

The relations \eqref{1.5} are of interest to us for two reasons. The first is that they are particularly useful when dealing with {\it sparse potentials}, which we will consider here. Sparse potentials are non-zero only at sites $x=x_n$ such that $x_n-x_{n-1}\to\infty$. This is because on the gaps --- the intervals where the potential is zero --- the propagation of $R$ and $\tht$ is especially transparent. Namely, $R$ is constant and $\tht$ increments by $k$ when passing from $x$ to $x+1$.

The second reason is that \eqref{1.5} provides a good control of the growth of $R$, which is the same as the growth of $u$ in the sense of the following lemma. Let us define 
\[
\|u\|_L^2\equiv \sum_{x=1}^L u(x)^2.
\]
Then we have

\begin{lemma} \lb{L1.1}
There are constants $c_1,c_2>0$ depending only on $k\in(0,\pi)$ such that if $u\sim(R,\tht)$ is any generalized eigenfunction for energy $2\cos(k)$ and $L\ge 2$, then
\[
c_1 \|u\|_L\le \|R\|_L \le c_2\|u\|_L.
\]
\end{lemma}

\smallskip
\noindent {\it Remark.}
The proof shows that $c_1$ and $c_2$ can be chosen uniformly for $k\in I$, with $I$ any closed sub-interval of $(0,\pi)$.
\smallskip

\noindent {\it Proof.}
From \eqref{1.4} we have
\begin{align*}
\allowdisplaybreaks
R(n)^2 &=u(n)^2+u(n-1)^2-2\cos(k) u(n)u(n-1)
\\ &\in \Bigl[ d_1\left(u(n)^2+u(n-1)^2\right), d_2\left(u(n)^2+u(n-1)^2\right) \Bigr]
\end{align*}
with $d_i=1+(-1)^i\cos(k)$. Hence
\[
d_1 \|u\|^2_L\le \|R\|^2_L \le 2d_2\bigl(\|u\|^2_L+u(0)^2\bigr).
\]
The result follows from the fact that
\[
u(0)^2=\big[(2\cos(k)-V(1))u(1)-u(2)\big]^2\le \left[(2\cos(k)-V(1))^2+1\right]\left[u(1)^2+u(2)^2\right]. \hfill\hfill\qed
\]
\smallskip

Let us denote by $u_{\phi,k}\sim(R_{\phi,k},\tht_{\phi,k})$ the generalized eigenfunction for energy $E=2\cos(k)$ satisfying the boundary condition $\phi$. We are now ready to state, in terms of $R$ rather than $u$, the abovementioned results from \cite{JL}. These will be our main tools for proving fractional dimension of measures.

\begin{proposition}[\cite{JL}] \lb{P1.2}
Let $0<\alpha\le 1$ and let $A$ be a Borel set of energies. If for every $E\in A$ and every generalized eigenfunction $u\sim(R,\tht)$ for energy $E$
\[
\lims_{L\to\infty}\frac{\|R\|_L^2}{L^{2-\alpha}} <\infty,
\]
then for any $\phi$ the restriction $\mu_{\phi}(A\cap\cdot)$ is $\alpha$-continuous.
\end{proposition}

This says that if all eigenfunctions for all energies in some support of $\mu_{\phi}$ have a small power growth, then $\mu_{\phi}$ cannot be very singular.

\begin{proposition}[\cite{JL}] \lb{P1.3}
Let $0<\alpha\le 1$ and let $A$ be a Borel set of energies. If for every $E\in A$  
\[
\limi_{L\to\infty}\frac{\|R_{\phi,k}\|_L^2}{L^\alpha}=0,
\]
where $k$ is such that $E=2\cos(k)$, then the restriction $\mu_{\phi}(A\cap\cdot)$ is $\alpha$-singular.
\end{proposition}

An eigenfunction $v$ for energy $E$ is called a {\it subordinate solution\/} if 
\[
\lim_{L\to\infty}\frac{\|v\|_L}{\|u\|_L}=0
\]
for any other eigenfunction $u$ with the same energy. The Gilbert-Pearson subordinacy theory \cite{GP} shows that $\mu_{\phi}$ is supported off the set of energies for which a subordinate solution exists, but does not satisfy the boundary condition $\phi$. Hence {\it Proposition \ref{P1.3}\/} says that the existence of a power decaying eigenfunction (which is then by standard arguments the subordinate solution) for all energies in some support of $\mu_{\phi}$ implies certain singularity of $\mu_{\phi}$. 

Moreover, {\it Lemma \ref{L2.1}\/} below shows that the existence of a power growing eigenfunction $u$ implies (for the potential \eqref{1.6}) the existence of a power decaying subordinate solution $v$, and the power of decay of $v$ is the same as the power of growth of $u$. Thus we only need to estimate the power of {\it growth} of eigenfunctions.

We will specifically concentrate on the generalized eigenfunctions with Dirichlet boundary condition $\phi=0$. Let us denote by $u_{k}$ the eigenfunction for energy $E=2\cos(k)$ with $k\in (0,\pi)$, such that $u_k(0)=0$ and $u_k(1)=1$. Let $u_k\sim(R_{k},\tht_{k})$, and let $\thh_{k}(x)=\tht_k(x)+k$. Notice that $R_k(1)=1$. Recall that $R_k(x)$, $\tht_k(x)$ and $\thh_k(x)$ are $C^\infty$ functions of $k$.

As we mentioned before, we will use sparse potentials, which are non-zero only for $x\in\{x_n\}_{n=1}^\infty$. It turns out that the set of possible candidates is quite small. Firstly, for our considerations we will need to have a good control of $\park\thh_{k}(x_n)$, in order to estimate the long-run behavior of $R_k$ using \eqref{1.5}. This turns out to be hard if $\{x_n\}_{n=1}^\infty$ grows sub-geometrically and so we will not consider this case. On the other hand we have

\begin{theorem} \lb{T1.4}
Let $x_n\in\bbN$ be an increasing sequence, $a_n\in\bbR$ and let $\mu_{\phi}$ be the spectral measure for the discrete Schr\" odinger operator on $\bbZ^+$ with boundary condition $\phi$ and potential
\[
V(x)=\begin{cases} a_n & \text{$x=x_n$ for some $n$},
\\ 0 & \text{otherwise}. \end{cases}
\]
Then
\begin{enumerate}
\item[{\rm(1)}] if $\lims x_n/x_{n+1}<1$ and $a_n\to 0$, then the dimension of $\mu_{\phi}$ is $1$ everywhere in $(-2,2)$;
\item[{\rm(2)}] if $x_n/x_{n+1}\to 0$ and $\sup |a_n|<\infty$, then the dimension of $\mu_{\phi}$ is $1$ everywhere in $(-2,2)$.
\end{enumerate}
\end{theorem}

\smallskip
\noindent {\it Remark.}
It is known \cite{KLS,KR} that in (2) the type of the spectrum depends on $\sum_{n=1}^\infty a_n^2$. If this is finite, then we have purely a.c.~spectrum, whereas if it is infinite, we have purely s.c.~spectrum.
\smallskip

\begin{proof}
(1) Let $I\subset(0,\pi)$ be a closed interval. For $k\in I$ let $w_k\sim(P_k,\psi_k)$ be the generalized eigenfunction with energy $E=2\cos(k)$ such that $w_k(0)=1$ and $w_k(1)=0$. 

There are $\ga>1$ and $n_0\in\bbN$ such that $x_n/x_{n-1}>\ga$ and $x_n>\ga^n$ for all $n>n_0$. Since $a_n\to 0$, for each $\eps>0$ there are $c_1,c_2>0$ such that $R_k^2(L)\ge c_1(1-\eps)^{n+1}$ and $P_k^2(L)\le c_2(1+\eps)^{n}$ for $L\in[x_{n}+1,x_{n+1}]$ and $k\in I$. This follows from \eqref{1.5}. Also, $L-x_{n-1}\ge c_0L$ with $c_0=1-\frac 1\ga>0$ if $n>n_0$. Let $\beta<1$. Then for $n>n_0$
\[
\frac{\|R_k\|_L^2}{\|P_k\|_L^{2\beta}}\ge\frac{(L-x_{n-1})c_1(1-\eps)^{n}}{\left(Lc_2(1+\eps)^{n}\right)^\beta}\ge cL^{1-\beta}\alpha^n\ge c\left(\ga^{1-\beta}\alpha\right)^n
\]
with $\alpha=(1-\eps)/(1+\eps)^\beta$. We can choose $\eps$ small enough so that $\ga^{1-\beta}\alpha>1$ and we obtain
\[
\lim_{L\to\infty} \frac{\|R_k\|_L^2}{\|P_k\|_L^{2\beta}}=\infty.
\]
Then {\it Theorem 1.2\/} from \cite{JL} implies $\tfrac{2\beta}{1+\beta}$-continuity of $\mu_0$. Since this is true for any $\beta<1$ and for any pair of generalized eigenfunctions with energy $E=2\cos(k)$, as well as for any $I$, the result follows.

(2) We know that $|a_n/\sin(k)|\le M<\infty$ for any $n$ and $k\in I$ ($I$ as above). It is easy to show that then there are $0<a<b<\infty$ such that
\[
1-\frac{a_n}{\sin(k)}\sin(2\tht)+\frac{a_n^2}{\sin^2(k)}\sin^2(\tht)\in (a,b)
\]
for any $n$, $\tht$ and $k\in I$. Therefore in the previous argument we have
to replace $1-\eps$ and $1+\eps$ with $a$ and $b$. On the other hand,
$x_n/x_{n+1}\to 0$ implies that for any $\ga$ there is $n_0$ such that
$x_n/x_{n-1}>\ga$ and $x_n>\ga^{n}$ for $n>n_0$. Thus for any $\beta<1$ choose $\ga$ large enough so that $\ga^{1-\beta}a/b^\beta>1$. The rest of the proof is identical with (1).
\end{proof}

This leaves us with non-decaying potentials and $x_n$ growing geometrically. Therefore we will consider the following natural choice of potential. We take $v\neq 0$ and an integer $\ga\ge 2$, and define
\begin{equation}\lb{1.6}
V_{v,\ga}(x)\equiv \begin{cases} v & \text{$x=x_n\equiv \ga^n$ for some $n\ge 1$},
\\ 0 & \text{otherwise}.\end{cases}
\end{equation}
Notice that the increasing gaps in $V_{v,\ga}$ ensure that the interval of
energies $[-2,2]$, which is the essential spectrum of
the free operator, is contained in the spectrum of each $H_{\phi}$. Indeed, by the discrete version of a theorem of Klaus ({\it Theorem
3.13\/} in \cite{CFKS}), the essential spectrum of $H_{\phi}$ is
\[
[-2,2]\cup\left\{ \sgn(v)\sqrt{4+v^2} \right\}.
\]
In what follows, we will show that for suitable $v$ and $\ga$,
the spectral measures $\mu_{\phi}$ are $\alpha$-continuous and $(1-\alpha)$-singular (for some $\alpha>0$) in some sub-interval of $[-2,2]$, and so have fractional
Hausdorff dimension there. It turns out that we will need small $|v|$ and large $\ga$.

More precisely, we will show, that for certain $v$ and $\ga$, we have that for ``most'' $k$ (in a sense to be specified later) within a given closed interval $I\subset (0,\pi)$ the function $R_{k}$ has a suitable power growth with some power $\beta\in(0,\frac 12)$. Then {\it Propositions \ref{P1.2}, \ref{P1.3}}, together with {\it Lemma \ref{L2.1}\/} below, can be used to compute (bounds on) the dimension of the spectral measures. Notice that since by \eqref{1.5} $R_{k}$ is constant on $[x_{n-1}+1,x_n]$, we only need to look at the growth of $\{R_{k}(x_n+1)\}_{n=1}^\infty$.

In what follows, we will not only prove fractional Hausdorff dimension, but also give bounds on it, and therefore we will need to enumerate those constants appearing in our argument, which affect the power of growth of $R_{k}$. We will denote these $C_i$.

In the present paper, we will consider spectral measures w.r.t.~$k$, rather than w.r.t.~$E=2\cos(k)$. This will not affect the validity of the results because on any closed interval $I\subset(0,\pi)$ of $k$'s (and we consider only such) the function $2\cos(k)$ is $C^1$ with bounded non-zero derivative. Therefore the dimensional properties of $\mu_{\phi}$ w.r.t.~$k\in I$ and w.r.t.~$E\in J\equiv 2\cos(I)$ are identical. Nevertheless, our results will be stated in terms of $E$.

Finally, we mention that in \eqref{1.6} $\ga$ does not need to be integral. If one only requires $\ga>1$ and sets, for instance, $x_n=\lfloor\ga^n\rfloor$, all our results continue to hold.

The rest of the paper is organized as follows. In Section \ref{S2} we present the main ideas of our proofs and results. Section \ref{S3} contains the abovementioned estimates on $\park\thh_k(x_n)$. In Section \ref{S4} we prove fractional Hausdorff dimension of the spectral measures for almost all boundary conditions, along with bounds on this dimension ({\it Theorems \ref{T4.1}, \ref{T4.2}\/}). Section \ref{S5} contains the same results for all boundary conditions ({\it Theorems \ref{T5.1}, \ref{T5.2}\/}). We distinguish these two cases because in the case of a.e.~$\phi$ the bounds we provide are considerably better than those for all $\phi$. In Section \ref{S6} we consider certain randomization of the potential $V_{v,\ga}$ given by \eqref{6.2} and prove exact fractional Hausdorff dimension for a.e.~realization of the potential and a.e.~boundary condition ({\it Theorem \ref{T6.3}}). Finally, Section \ref{S7} contains the corresponding whole-line results ({\it Theorem \ref{T7.1}}). 

It is a pleasure to thank Barry Simon for many useful discussions and suggestions. My thanks also go to David Damanik, Wilhelm Schlag and Boris Solomyak.


\section{Growth of Eigenfunctions} \lb{S2}

In this section, we will present a short tour of the proof of our main result. Technical details are left for later.

Let $J\subset(-2,2)$ be a given closed interval of energies. We let $I\subset(0,\pi)$ be such that $2\cos(I)=J$. We define $v_k\equiv -\frac v{\sin(k)}$ for $k\in I$, $w_1\equiv \min_{k\in I}\{|v_k|\}>0$, $w_2\equiv \max_{k\in I}\{|v_k|\}<\infty$, $x_n\equiv \ga^n$ for $n\ge 1$ and $x_0\equiv 0$. Here $v\neq 0$ and $\ga\ge 2$ are to be determined later. We consider the half-line discrete Schr\" odinger operator $H_{\phi}$ given by \eqref{1.1}, \eqref{1.2} with the potential \eqref{1.6}, and we denote its spectral measure $\mu_{\phi}$. We will prove fractional Hausdorff dimension of $\mu_{\phi}(I\cap\cdot)$ for suitable $v$, $\ga$ and $\phi$.

As mentioned earlier, we need to estimate the growth of $R_k(x_n+1)$. As in \cite{KLS}, using \eqref{1.5} and the Taylor series of $\ln(1+x)$, one obtains for $n\ge 1$
\begin{align} \lb{2.1}
\allowdisplaybreaks
\ln \bigl(R_k(x_n+1)\bigr)&-\ln \bigl(R_k(x_{n-1}+1)\bigr) = \frac 12\ln\Bigl(1+v_k\sin(2\thh_k(x_n)) + v_k^2\sin^2(\thh_k(x_n))\Bigr) \notag
\\ &= \frac 12 v_k\sin(2\thh_k(x_n)) + \frac 14 v_k^2\Bigl(2\sin^2(\thh_k(x_n)) - \sin^2(2\thh_k(x_n))\Bigr) + g_n(k) \notag
\\ &= \frac{v_k^2}8 + \frac{v_k}2\sin(2\thh_k(x_n)) + \frac{v_k^2}8\Bigl(\cos(4\thh_k(x_n))-2\cos(2\thh_k(x_n))\Bigr) + g_n(k)
\end{align}
where $|g_n(k)|<C_0|v_k|^3$ for some $C_0>0$. Here $g_n(k)$ is the sum of all third and higher order terms in $v_k$, and the last equality comes from 
\[
2\sin^2(\tht)-\sin^2(2\tht)=\tfrac 12-\cos(2\tht)+\tfrac 12\cos(4\tht).
\]

Now take $v\neq 0$ small enough so that $w_2+w_2^2<1$ and $C_0w_2^3<
w_1^2/8$, and define $d_1\equiv w_1^2/8-C_0w_2^3$ and
$d_2\equiv w_2^2/8+C_0w_2^3$ If we let $0<C_1<d_1$ and $d_2<C_2<\infty$, then we have
\begin{equation} \lb{2.2}
\frac{v_k^2}8+g_n(k)\in [d_1,d_2] \subset (C_1,C_2)
\end{equation}
for any $n$ and $k\in I$. From now on, $v$ and $C_0$ (and thus also $C_1$ and $C_2$) will be fixed.

Here is our main idea. It follows from \eqref{2.2} that the contribution of the terms $v_k^2/8+g_n(k)$ to the size of $\ln (R_k(x_n+1))$ is within the interval $[d_1n,d_2n]$. If we could show that for
large $n$ the contribution of the remaining three (oscillating) terms in \eqref{2.1} is
small compared to this, we would obtain estimates proving positive power of
the growth of $R_k$. It is reasonable to hope for this because the
oscillating terms change sign when $n$ is varied, and so we can expect
cancellations. This is the central idea of \cite{KLS}.

So let us assume for a while that for some $A\subseteq I$ with $|A|=0$ ($|A|$ being the Lebesgue measure of $A$)
\begin{equation} \lb{2.3}
\sum_{n=1}^N\sin(2\thh_k(x_n))=o(N) \qquad \forall k\in I\backslash A 
\end{equation}
and that the same holds for the other two oscillating terms. For any $k\in I\backslash A$ we
have for large $n$ (by \eqref{2.2})
\begin{equation} \lb{2.4}
R_k(x_n+1)\in(e^{C_1 n},e^{C_2 n})=(x_n^{\beta_1},x_n^{\beta_2}) 
\end{equation}
where $\beta_i\equiv C_i/\ln(\ga)$. If we choose $\ga$ to be large enough so that $0<\beta_1<\beta_2<\frac 12$, we get 
\begin{equation} \lb{2.5}
c_1 L^{1+2\beta_1}\le\|R_k\|_L^2\le c_2L^{1+2\beta_2}
\end{equation}
for large $L$ and $c_i=c_i(k)$. Then {\it Proposition \ref{P1.2}\/}, along with the theory of rank one perturbations \cite{Si}, proves for a.e.~boundary condition $\phi$ that the dimension of $\mu_{\phi}$ is at least $1-2\beta_2$ (see the last paragraph of the proof of {\it Theorem \ref{T4.1}}).

To obtain a good upper bound for the dimension, we need to prove an appropriate decay of the corresponding subordinate solutions. We will use the following result, the proof of which we postpone until the end of this section.

\begin{lemma} \lb{L2.1}
Let $x_1<x_2<\cdots$ be such that $\| T_E(x_n,x_{n-1})\|\le B$ for some $E\in(-2,2)$ and $B<\infty$. Let us assume that $u\sim (R,\tht)$ is a generalized eigenfunction for energy $E$ such that
\[
R(x_n+1)=e^{\alpha_n}
\]
where $\alpha_n=\sum_{1}^n (Z_j+X_j)$ with $Z_j\in[d_1,d_2]$ {\rm (}for some
$0<d_1\le d_2<\infty${\rm )} and $\sum_1^n X_j=o(n)$. Then there exists a
subordinate solution $v\sim(P,\psi)$ for energy $E$ such that for any $0<d<d_1$ and for all sufficiently large $n$ we have
\[
P(x_n+1)\le e^{-dn}.
\]
\end{lemma}

\smallskip
\noindent {\it Remark.} 
In principle, the lemma shows that the power of decay of the subordinate solution is the same as the power of growth of all other solutions for the same energy. Hence, in a sense, the result is optimal. Notice that information on only one growing solution is needed, but this has to satisfy a ``steady growth'' condition. Compare with Lemma 8.7 in \cite{KLS}, where two growing solutions are involved.
\smallskip

The lemma can be applied here because we have \eqref{2.2} and \eqref{2.3}.
Notice also that all the powers of the free transfer matrix
\[
\left(\begin{matrix} E&-1\\1&0\end{matrix}\right)
\]
for energy $E\in(-2,2)$ are uniformly norm-bounded. Hence, we also have a uniform
(in $n$) bound on
\[
\|T_E(x_n,x_{n-1})\|\le 
\Bigg\|\left(\begin{matrix} E-v&-1\\1&0\end{matrix}\right)\Bigg\| \cdot
\Bigg\|\left(\begin{matrix} E&-1\\1&0\end{matrix}\right)^{x_n-x_{n-1}-1}\Bigg\|.
\]
So for $k\in I\backslash A$ we let
\[
Z_j=\frac{v_k^2}8+g_j(k)
\]
and
\[
X_j=\frac{v_k}2\sin(2\thh_k(x_j)) +
\frac{v_k^2}8\Bigl(\cos(4\thh_k(x_j))-2\cos(2\thh_k(x_j))\Bigr).
\]
If we now take $d$ such that $C_1<d<d_1$ (say $d=C_1+\eps$), we obtain the
existence of a vector $\overline{u}_k^{\,\text{sub}}\in\bbR^2$ which generates the
subordinate solution $u_k^{\text{sub}}$ with $\|u_k^{\text{sub}}\|^2_L\le
L^{1-2\beta_1-2\eps/\ln(\ga)}$.
Hence by {\it Proposition \ref{P1.3}\/} and the Gilbert-Pearson subordinacy theory \cite{GP}, the dimension of $\mu_{\phi}$
is at most $1-2\beta_1$ for a.e.~$\phi$ (see the proof of {\it Theorem \ref{T4.1}}).

Moreover, by the proof of {\it Lemma 2.1}, $\| T_E(x_n)\| \to\infty$ as $n\to\infty$ whenever $E\in 2\cos(I\backslash A)$. {\it Theorem 1.2\/} from \cite{LS} then implies absence of a.c.~spectrum in $I$ for all $\phi$. The next paragraph proves absence of p.p.~spectrum if $w_2+w_2^2<\ln(\ga)$. Thus for $\ga$ large enough, we obtain purely s.c.~spectrum in $I$ for all $\phi$.

But we can do even better. It turns out that with a slightly stronger assumption than \eqref{2.3}, we can show fractional dimension {\it for all\/} boundary conditions! First notice that 
\[
\ln \bigl( R_k(x_n+1) \bigr) -\ln \bigl( R_k(x_{n-1}+1) \bigr) < \frac{w_2+w_2^2}2
\]
and so for all $k\in I$ we have $R_k(x_n+1)<x_n^{(w_2+w_2^2)/2\ln(\ga)}$. Since $R_k$ is constant on $[x_{n-1}+1,x_n]$, {\it Proposition \ref{P1.2}\/} implies that the dimension is at least $1-(w_2+w_2^2)/\ln(\ga)$ on all of $I$. Since $w_2+w_2^2<1$,  this is strictly positive if $\ga\ge 3$.

Now replace \eqref{2.3} with the (stronger) assumption that for any
$\eps>0$ there exists $A^\eps\subseteq I$, a set with dimension smaller than 1
such that
\begin{equation} \lb{2.6}
\lims_{N\to\infty} \frac 1N\left|\sum_{n=1}^N\sin(2\thh_k(x_n))\right|\le\eps \qquad
\forall k\in I\backslash A^\eps,
\end{equation}
and the same is true for the other two oscillating terms. Let us consider the ``Dirichlet'' measure $\mu_0$. Since it has no a.c.~part, it is supported on the set of $k$ for which $u_k$ is the subordinate solution. It follows that $\mu_0(I\backslash A^\eps)=0$. Indeed, if $k\in I\backslash A^\eps$, then for large $n$ we have
\[
R_k(x_n+1) > \exp \biggl\{ \left(C_1-\eps \left( \frac{w_2}2 + \frac{3w_2^2}8 \right) \right) n \biggr\}.
\]
If $\eps$ is small enough so that the exponent is positive, {\it Lemma \ref{L2.1}\/} again shows the existence of a subordinate solution $u_k^{\text{sub}}$, and this must be different from $u_k$ (by which we mean that $u_k^{\text{sub}}$ is not a multiple of $u_k$). This implies that $\mu_0$ is supported on $A^\eps$.

Hence the dimension of $\mu_0$ on $I$ (which we know is positive) is at most $\dim(A^\eps)<1$, and therefore fractional. Considering instead of $u_k$ the generalized eigenfunction for energy $2\cos(k)$
satisfying boundary condition $\phi\neq 0$, and assuming \eqref{2.6} for the
corresponding $\thh_{\phi,k}(x_n)$, we can obtain the same result
for any $\phi$. For details see the proof of {\it Theorem \ref{T5.1}}.

Our considerations have, however, a pitfall. Neither \eqref{2.3} nor
\eqref{2.6} need hold for such a large set of $k$'s as we want. We
cannot hope for this because we have only limited control of
the argument of the $\sin$ term. Fortunately, we do not really need
\eqref{2.3} and \eqref{2.6} in the presented form. This is because the
non-oscillatory term $v_k^2/8$ gives us some space. We can
``sacrifice'' part of it, just as we did when we joined it with the $g_n(k)$
term, and still keep a power growth of $R_k$. More precisely, we will divide 
the $\sin$ term into two, one of which will be
small with respect to the non-oscillatory term, whereas the other one will
have enough ``regularity'' for \eqref{2.3} and \eqref{2.6} to hold. The two
$\cos$ terms can be treated similarly.

\begin{proof}[Proof of Lemma~\ref{L2.1}]
Let $d_0=\tfrac 12(d+d_1)$ and let $v\sim(P,\psi)$ be any solution for energy $E$ different
from $u$. Let $p_n\equiv P(x_n+1)$, $r_n\equiv R(x_n+1)$ and $t_n\equiv \|T_E(x_n)\|$.
The first ingredient in the proof is {\it Theorem 2.3\/} from \cite{KLS}
which states that there are $c_1,c_2>0$ such that
\[
c_1\max\{p_n,r_n\}\le t_n\le c_2\max\{p_n,r_n\}.
\]

Since by hypothesis $r_n\ge e^{d_0n}/c_1$ for large $n$, it follows that for large enough $n$
\begin{equation} \lb{2.7}
t_n\ge e^{d_0n}.
\end{equation}
On the other hand, if $d_3>\max\{ d_2,\ln(B) \}$,
then for large $n$
\[
t_n\le e^{d_3n}.
\]
Hence if
\[
\delta\equiv \limi_{n\to\infty}\frac{\ln(t_n)}n,
\]
then $d_0\le \delta\le d_3$.

From \eqref{2.7} we know that 
\[
\sum_{n=1}^\infty \frac{\|T_E(x_n,x_{n-1})\|^{2}} {\|T_E(x_n)\|^{2}} <\infty.
\]
This is the assumption of
our second ingredient, {\it Theorem 8.1\/} from \cite{LS}, the proof of
which (namely inequalities (8.5), (8.7)) yields the following.
There is a vector $\overline{v}\in\bbR^2$ and $c_0>0$ such that
\[
\|T_E(x_n)\overline{v}\|^2\le c_0t_n^{2}s_n^2+t_n^{-2}
\]
with $s_n=\sum_{m=0}^\infty t_{n+m}^{-2}$.
Our aim is to show that for large enough $n$ this is smaller than
$e^{-2dn}$.

The abovementioned {\it Theorem 8.1\/} also asserts that $\overline{v}$
generates the subordinate solution for energy $E$. One expects that this is
different from $u$, generated by $\overline{u}\equiv (u(0),u(1))$,
which is a growing solution. We will now prove this claim.

Let $\delta'<\delta<\delta''$ be such that $\delta''<2\delta'$.
From the definition of $\delta$ we know that for large enough $n$
\[
s_n\le e^{-2\delta'n}.
\]
Also there are $\{n_j\}_{j=1}^\infty$ such that $t_{n_j}<e^{\delta''n_j}$ and
so
\[
\|T_E(x_{n_j})\overline{v}\|^2\le c_0e^{2\delta''n_j}e^{-4\delta'n_j}+
e^{-2\delta''n_j}\to 0
\]
as $j\to\infty$. Since
\[
\|T_E(x_{n_j})\overline{u}\|^2=u(x_{n_j}+1)^2+u(x_{n_j})^2\ge
\frac{R(x_{n_j}+1)^2}2\to\infty,
\]
the subordinate solution is indeed different from $u$. Let us take it for $v$, and change $P$, $p_n$, $c_1$ and $c_2$
accordingly. Since {\it Theorem 8.1\/} from \cite{LS} also states that
\[
\frac{\|T_E(x_n)\overline{v}\|}{\|T_E(x_n)\overline{u}\|}\to 0,
\]
it then follows that $p_n<r_n$ for large $n$. Hence $c_1r_n<t_n<c_2r_n$ for large $n$.

Let $a_n=\alpha_n/n$. If $n$ is large enough, then $a_n\in(d_0,d_3)$. Let
$b\in\bbN$ be such that $d_0(b+1)>d_3$. Pick $\eps>0$ small enough so
that $\eps'\equiv \eps(b+2)<(d_0-d)/2$.

Since $a_nn=\sum_1^n (Z_j+X_j)$, for large $n$ we have $\sum_1^n Z_j\ge
(a_n-\eps)n$ and $|\sum_1^n X_j|\le \eps n$. Since
\[
\ln(r_{n+m})=a_{n+m}(n+m)=\sum_1^{n+m}(Z_j+X_j)\ge
(a_n-\eps)n+d_1m-\eps(n+m),
\]
for $m\le bn$ we have
\[
r_{n+m}\ge e^{(a_n-\eps')n+d_1m}.
\]
Hence for large $n$
\[
s_n\le c_1^{-2}\sum_{m=0}^{bn} e^{-2(a_n-\eps')n-2d_1m}+c_1^{-2}
\sum_{m=bn}^\infty
e^{-2d_0(n+m)}\le c\left(e^{-2(a_n-\eps')n}+e^{-2d_0(b+1)n}\right)
\]
for some $c>0$.
It follows that
\begin{align*}
\|T_E(x_n)\overline{v}\|^2
& \le 2cc_0c_2^2\left(e^{(2a_n-4(a_n-\eps'))n}+e^{(2a_n-4d_0(b+1))n}\right)
+c_1^{-2}e^{-2a_nn}
\\ & \le c'\left(e^{-2dn}+e^{-2d_3n}+e^{-2d_0n}\right) \, \le  \, 3c'e^{-2dn}.
\end{align*}

The rest is an easy computation. From \eqref{1.4} we have
\[
P^2(x_n+1)\le 2\left( v(x_n+1)^2+v(x_n)^2\right)
=2\|T_E(x_n)\overline{v}\|^2\le 6c'e^{-2dn}
\]
for large $n$. Since this holds for any $d<d_1$, the result follows.
\end{proof}


\section{Estimates on Growth of $\thh$} \lb{S3}

From now on let us write $\thh(x_n,k)$ instead of $\thh_k(x_n)$. 
As mentioned before, the key to our results are estimates on $\park\thh(x_n,k)$ which we
denote $\thh'(x_n,k)$. Differentiating the first equation of \eqref{1.5} with respect to $k$, and using $\sin^2(\tht)=(1+\cot^2(\tht))^{-1}$, we get (as in \cite{KR})
\[
\park\tht(x+1)=\frac{\park\thh(x)-V(x)\frac{\cos(k)}{\sin^2(k)}\sin^2(\thh(x))} {\sin^2(\thh(x))+\left[\cos(\thh(x))-\frac{V(x)}{\sin(k)}\sin(\thh(x))\right]^2}.
\]
For our potential $V_{v,\ga}$ and for $x=x_n$, the denominator is 
\[
a_{n,k}=1+v_k\sin(2\thh(x_n,k))+v_k^2\sin^2(\thh(x_n,k))
\]
and so
\begin{equation} \lb{3.1}
\begin{aligned}
\allowdisplaybreaks
\thh'(x_{n+1},k) &=x_{n+1}-x_n+\tht'(x_n+1,k)
\\ &=x_{n+1}-x_n + \frac{\thh'(x_n,k)}{a_{n,k}} + \frac{v_k\cot(k)\sin^2(\thh(x_n,k))}{a_{n,k}}. 
\end{aligned}
\end{equation}

The denominator $a_{n,k}$ is in the interval $[C_3,C_4]\subset(1-(w_2+w_2^2),1+(w_2+w_2^2))$ with 
\[
C_i\equiv 1+\frac{w_2^2}2+(-1)^iw_2\sqrt{1+\frac{w_2^2}4}.
\]
Notice that $C_3C_4=1$. 
The last fraction in \eqref{3.1} is in some interval $(-M,M)$ for any $n$
and any $k\in I$. From \eqref{3.1} one can show that as $n\to\infty$,
$\thh'(x_n,k)$ gets close to $x_n$. Indeed, if for some $n$ we have
$\thh'(x_n,k)/x_n\in(\Delta_1,\Delta_2)$ for all $k\in I$ (and some $\Delta_i\in\bbR$), then
\[
\frac{\thh'(x_{n+1},k)}{x_{n+1}} \in \left( 1 + \frac 1\ga \left( \frac{\Delta_1}{C_4}-1 \right) 
- \frac M{\ga^{n+1}} ,\, 1 + \frac 1\ga \left( \frac{\Delta_2}{C_3}-1 \right) + \frac M{\ga^{n+1}} \right).
\]
Iterating this, one proves that for any $D_1<C_5\equiv (\ga-1)/(\ga-C_3)$ and
$D_2>C_6\equiv (\ga-1)/(\ga-C_4)$ there is $n_0$ such that for $n>n_0$ and all $k\in I$
\begin{equation} \lb{3.2}
\frac{\thh'(x_n,k)}{x_n}\in(D_1,D_2).
\end{equation}
This is because $(\ga-1)/(\ga-C_i^{-1})$ are the fixed points of the map $\Delta\mapsto 1+(\Delta/C_i-1)\ga^{-1}$, and $C_3^{-1}=C_4$. Thus $(D_1,D_2)$ is an interval containing 1, which can be made as small as we need by taking $\ga$ large enough. This will play an important role in our considerations.

For $n>n_0$ let $K_{n,1}$ and $K_{n,2}$ be the smallest and largest numbers in
$I$ such that $\thh(x_n,K_{n,i})$ is an integral multiple of $\pi$. Notice
that by \eqref{3.2} the distance of these numbers from the corresponding
edges of $I$ is smaller than $\pi D_1^{-1}\ga^{-n}$. Let
$k_{n,1}=K_{n,1}<k_{n,2}<\dots<k_{n,j_n}=K_{n,2}$ be all numbers in $I$
such that $\thh(x_n,k_j)$ is an integral multiple of $\pi$.
Let $I_{n,j}=[k_{n,j},k_{n,j+1}]$. Then \eqref{3.2} implies that
\begin{equation} \lb{3.3}
|I_{n,j}|\in(\pi D_2^{-1}\ga^{-n},\pi D_1^{-1}\ga^{-n}).
\end{equation}
We will slightly alter the oscillating $\sin$ term on each $I_{n,j}$. This way we will
obtain the (previously mentioned) more regular term for which we can prove \eqref{2.3} (and later \eqref{2.6}). This is the content of Sections \ref{S4} and \ref{S5}. But before that, we present an additional argument.

It turns out that by considering $\thh''(x_n,k)$ one can improve \eqref{3.2} and \eqref{3.3} on small scale (i.e., the scale of $I_{n,j}$). This improvement is not essential for our results; it only yields better numerical estimates for large $\ga$. Differentiating \eqref{3.1} with respect to $k$ one obtains
\[
\thh''(x_{n+1},k)=\frac{\thh''(x_n,k)}{a_{n,k}} - \Bigl[ \thh'(x_n,k) \Bigr]^2 \frac{2v_k\cos(2\thh(x_n,k))+v_k^2\sin(2\thh(x_n,k))}{a_{n,k}^2} + b_{n,k}\thh'(x_n,k) + \tilde b_{n,k}
\]
where $b_{n,k},\tilde b_{n,k}\in (-M,M)$ for some $M$ and all $n$, $k$. Then for $n>n_0$ (since $a_{n,k}^{-1}\le C_4$),
\[
\frac{|\thh''(x_{n+1},k)|}{x_{n+1}^2}\le \frac{C_4}{\ga^2}\frac{|\thh''(x_{n},k)|}{x_{n}^2}+ \frac{D_2^2C_4^2(2w_2+w_2^2)}{\ga^2}+\frac{2MD_2}{\ga^{n+1}}.
\]
Similarly as above, by iterating this we obtain that for large enough $n$ (say $n>n_0$ for a new $n_0$) and all $k\in I$
\[
\frac{|\thh''(x_{n},k)|}{x_{n}^2}\le \frac{D_2^2C_4^2(2w_2+w_2^2)}{\ga^2-2}.
\]
This is because the fixed point of the mapping $\Delta\mapsto C_4\Delta/\ga^2+D_2^2C_4^2(2w_2+w_2^2)/\ga^2$ is $D_2^2C_4^2(2w_2+w_2^2)/(\ga^2-C_4)$, and because $C_4<2$.

If $n>n_0$, let $D_1^{n,j}$ and $D_2^{n,j}$ be such that for any $k\in I_{n,j}$
\begin{equation} \tag{\ref{3.2}$'$}
\frac{\thh'(x_n,k)}{x_n}\in [D_1^{n,j},D_2^{n,j}],
\end{equation}
and the interval $[D_1^{n,j},D_2^{n,j}]$ is smallest possible. From \eqref{3.3} and the obtained estimate on $|\thh''(x_n,k)|$ we have that
\[
D_2^{n,j}-D_1^{n,j} \le \frac \pi{D_1\ga^{n}} \frac{D_2^2C_4^2(2w_2+w_2^2)}{\ga^2-2} \ga^{2n}\ga^{-n}=\frac{\pi D_2^2C_4^2(2w_2+w_2^2)}{D_1(\ga^2-2)}.
\]
Notice that $D_2-D_1\approx c/\ga$, and so as $\ga\to\infty$, the estimate (\ref{3.2}$'$) is better than \eqref{3.2}. We also have the obvious improvement of \eqref{3.3}, namely
\begin{equation} \tag{\ref{3.3}$'$}
|I_{n,j}|\in\left[\pi (D_2^{n,j})^{-1}\ga^{-n},\pi (D_1^{n,j})^{-1}\ga^{-n}\right].
\end{equation} 

Now let
\[
C_7\equiv \min\left \{\frac{\pi D_2^2C_4^2(2w_2+w_2^2)}{D_1^2(\ga^2-2)}, \frac{D_2-D_1}{D_1} \right\}.
\]
Since $D_1\le D_1^{n,j}\le D_2^{n,j}\le D_2$, we have
\[
\frac{D_2^{n,j}-D_1^{n,j}}{D_1^{n,j}}\le C_7 \le \frac{D_2-D_1}{D_1}
\]
for all $j$ and $n>n_0$.


\section{Fractional Dimension for A.E.~Boundary Condition}  \lb{S4}

Let us now turn to the announced division of the $\sin$ term in two. 
Since $(C_6-C_5)/C_5=(C_4-C_3)/(\ga-C_4)$, we can pick $D_1$, $D_2$ close to $C_5$, $C_6$ so
that $(D_2-D_1)/D_1$ is arbitrarily close to $(C_4-C_3)/(\ga-C_4)$. Since
$C_3$ and $C_4$ (as well as $C_1$ and $C_2$) are independent of $\ga$, and
$C_7\le (D_2-D_1)/D_1$, we can make $C_7$ arbitrarily small by taking $\ga$
large enough. Let us do this so that
\begin{equation} \lb{4.1}
\frac{C_7}\pi \left(\frac{w_2}2+\frac{3w_2^2}8\right)<C_1.
\end{equation}

For $k\in I$ and $n>n_0$ we let
\[
Q_n(k)\equiv \begin{cases} |I_{n,j}|^{-1} \int_{I_{n,j}} \sin(2\thh(x_n,\kappa)) \,d\kappa & k\in I_{n,j} \\ 0 & k\in I\backslash [K_{n,1},K_{n,2}] \end{cases}
\]
and define
\[
X_n(k)\equiv \sin(2\thh(x_n,k))-Q_n(k).
\]
Notice that
\begin{equation} \lb{4.2}
\int_{I_{n,j}}X_n(k)\,dk=0
\end{equation}
for any $j$. Also, by (\ref{3.2}$'$)
\[
|\sin(2\thh(x_n,k))-X_n(k)|=|Q_n(k)|\le \frac 2\pi \frac{D_1^{n,j}-D_2^{n,j}}{D_1^{n,j}+D_2^{n,j}}\le \frac {C_7}\pi
\]
for any $k\in I$ and $n>n_0$.
This is because $|Q_n|$ is maximal possible on $I_{n,j}$ if $\thh'$ equals $D_1^{n,j}\ga^n$ when $\sin(2\thh)$ is positive and $D_2^{n,j}\ga^n$ when $\sin(2\thh)$ is negative (or vice-versa). And in that case we have equality in the first inequality above. 

Therefore we have
\[
\left|\sum_{n=n_0+1}^N\frac{v_k}2\sin(2\thh(x_n,k))\right|\le
\frac{w_2}2 \frac{C_7}\pi N + \left|\frac{v_k}2\sum_{n=n_0+1}^N X_n(k)\right|.
\]
If we do the same with the other two oscillating terms (the corresponding $I_{n,j}$'s and $X_n$'s will be slightly different), the three terms containing $C_7$ will add up to 
\[
\frac{C_7}\pi\left(\frac{w_2}2+\frac{3w_2^2}8\right)N < C_1N.
\]
So if we prove \eqref{2.3} for $X_n$ in place of $\sin$ (and similarly for the two $\cos$ terms), we will still keep a power growth of $R_k$. We will be able to do this using \eqref{4.2}.

To this end we need estimates on the covariances of the $X_n$'s, so take $n>m>n_0$. Then for all $j$ we have \eqref{4.2}, whereas from \eqref{3.2} it follows that $X_m$ takes on $I_{n,j}$ values within an interval of length at most $D_2\ga^m|I_{n,j}|\le \pi D_2D_1^{-1}\ga^{m-n}$. Therefore
\[
\left|\int_{I_{n,j}} X_mX_n\,dk\right|\le \frac{\pi D_2}{D_1}\ga^{m-n}|I_{n,j}|.
\]
Notice that this might not be true if $I_{n,j}$ contains some $k_{m,l}$, points where $X_m$ may be discontinuous. But this is not a problem because the total length of such $I_{n,j}$'s is of order $\ga^{m-n}$. Since also $|I\backslash [K_{n,1},K_{n,2}]|\sim \ga^{-n}$ and the $X_n$'s are bounded, we get for $n>m>n_0$
\[
\left|\int_I X_mX_n\,dk\right|\le c\ga^{m-n}.
\]

By Cauchy-Schwarz inequality
\begin{align*}
\allowdisplaybreaks
\int_I\left|\sum_{n=n_0+1}^N X_n\right|\,dk 
& \le c \left[ \int_I\left(\sum_{n=n_0+1}^N X_n\right)^2\,dk \right]^{1/2}
\\ & = c \left[ \int_I\sum_{n=n_0+1}^N X_n^2+\sum_{n_0< m<n\le N}2X_mX_n\,dk \right]^{1/2} 
\le \, c\left( N+cN \right)^{1/2} \, \le \, cN^{1/2}
\end{align*}
(the value of $c$ changes in each inequality, but is independent of $N$). Thus
\[
\int_I\left|\frac{\sum_{n=n_0+1}^{N^4} X_n}{N^4}\right|\,dk\le \frac
c{N^2}
\]
which means that for a.e.~$k$
\[
\frac{\sum_{n=n_0+1}^{N^4} X_n}{N^4}\to 0,
\]
because $\sum_1^\infty N^{-2}<\infty$.
Since the $X_n$'s are bounded and $[(N+1)^4-N^4]/N^4\to 0$, it follows that for these $k$
\[
\frac{\sum_{n=n_0+1}^{N} X_n}{N}\to 0
\]
and so
\[
\frac{\sum_{n=1}^{N} X_n}{N}\to 0.
\]

\begin{theorem} \lb{T4.1}
Let $H_{\phi}$ be the discrete Schr\" odinger operator on $\bbZ^+$ with potential $V_{v,\ga}$ given by \eqref{1.6}, and boundary condition $\phi$. Let $\mu_{\phi}$ be its spectral measure. For any closed interval of energies $J\subset(-2,2)$ there is $v_0>0$ and $\ga_0\in\bbN$ such that if $0<|v|<v_0$ and $\ga\ge\ga_0$ is an integer, then each $\mu_{\phi}$ is purely singular continuous in $J$, and for a.e.~$\phi$ the measure $\mu_{\phi}$ has fractional Hausdorff dimension in $J$.
\end{theorem}

\smallskip
\noindent {\it Remark.}
A priori, $\ga_0$ depends on $v$. However, since in \eqref{4.1} we have $C_1=O(v^2)$, $C_7=O(v\ga^{-2})$ and $w_2=O(v)$ as $v\to 0$, $\ga\to\infty$, we can choose $\ga_0$ uniformly for all small $v$.
\smallskip

\begin{proof}
Let $I\subset(0,\pi)$ be such that $2\cos(I)=J$. The above discussion and \eqref{2.2} show that there is $A\subseteq I$ with $|A|=0$, such that for any $k\in I\backslash A$ there is $n_1(k)$ such that for $n>n_1(k)$
\[
R_{k}(x_n+1)\in(e^{c_1 n},e^{c_2 n})=(x_n^{\beta_1},x_n^{\beta_2}),
\]
where 
\[
c_i\equiv C_i+(-1)^i\frac{C_7}\pi \left(\frac{w_2}2+\frac{3w_2^2}8\right) > 0 \qquad (i=1,2)
\]
and $\beta_i\equiv c_i/\ln(\ga)$. Take $\ga$ large enough so that $0<\beta_1<\beta_2<\frac 12$.
It follows from the constancy of $R_{k}$ on $[x_n+1,x_{n+1}]$ that \eqref{2.5} holds for $k\in I\backslash A$.

Using {\it Lemma \ref{L2.1}\/} with 
\[
d_i+(-1)^i\frac{C_7}\pi \left(\frac{w_2}2+\frac{3w_2^2}8\right) \qquad (i=1,2)
\]
in place of $d_1$, $d_2$, one gets for these $k$ the existence
of a vector $\overline{u}_k^{\,\text{sub}}\in\bbR^2$ such that the following holds. If $u_k^{\text{sub}}\sim(P_k,\psi_k)$ is the generalized eigenfunction for energy $2\cos(k)$ generated by $\overline{u}_k^{\,\text{sub}}$, then for some small $\eps$
\[
P_k(x_n+1)\le e^{-(c_1+\eps)n}=x_n^{-\beta_1-\eps/\ln(\ga)}.
\]
Since $P_k$ is constant on $[x_{n-1}+1,x_n]$, we have
\begin{equation} \lb{4.3}
\|P_k\|^2_L\le c'L^{1-2\beta_1-2\eps/\ln(\ga)}.
\end{equation}

Since $u^{\text{sub}}_k$ is the subordinate solution for energy $2\cos(k)$, all
other solutions for this energy grow (in power) no faster than $R_{k}$.
Also, $\mu_{\phi}$ restricted to $I$ is supported on the set of those $k$, for which
$u_k^{\text{sub}}$ satisfies the boundary condition $\phi$ (this is because
$\mu_{\phi}$ has no a.c.~part; see \cite{GP}), and so $P_k=R_{\phi,k}$ . Then we have by 
{\it Propositions \ref{P1.2}\/}, {\it \ref{P1.3}\/} and by \eqref{2.5}, \eqref{4.3} that for any $\phi$
the restriction $\mu_{\phi}((I\backslash A)\cap\cdot)$ is
$(1-2\beta_2)$-continuous and $(1-2\beta_1)$-singular. By the theory of
rank one perturbations ({\it Theorem 1.8\/} in \cite{Si}) we know that
$\mu_{\phi}(A)=0$ for a.e.~$\phi$, and so for a.e.~$\phi$ we have the
same continuity/singularity of $\mu_{\phi}(I\cap\cdot)$. 
\end{proof}

To get numerical bounds on the dimensions, one needs to evaluate constants $C_i$. These depend on $w_i$ and therefore it is best to consider (for given $v$ and large $\ga$) a small interval $I$ around each $k$, so that $w_i\approx |v_k|$. One then obtains bounds on local dimension of $\mu_{\phi}$ which will be $k$-dependent. We will, however, first present an additional argument which will considerably facilitate this (by eliminating constants $C_0$, $C_1$ and $C_2$) as well as improve the obtained bounds.

Let us push the above ideas a little bit further. At the beginning of this argument we introduced the term $g_n(k)$ as the sum of all third and higher order terms in $v_k$. One can, however, write down all these terms explicitly, using the Taylor series of $\ln(1+x)$.
\begin{align*}
\allowdisplaybreaks
\ln \bigl(R_k(x_n+1)\bigr) & -\ln \bigl(R_k(x_{n-1}+1)\bigr) 
= \tfrac 12\ln\Bigl(1+v_k\sin(2\thh(x_n,k)) + v_k^2\sin^2(\thh(x_n,k))\Bigr)
\\ & = \sum_{\substack{a,b\ge 0 \\ a+b\ge 1}} \frac{(-1)^{a+b+1}}{2a+2b}
\binom{a+b}a v_k^{a+2b}\sin^{a}(2\thh(x_n,k))\sin^{2b}(\thh(x_n,k)).
\end{align*}
If $|v_k|+v_k^2<1$, then the sum of the amplitudes of the terms of this sum (with $n$ fixed) converges absolutely.

Now notice that all terms with odd $a$ are oscillating, whereas terms with even $a$ do not change sign when varying $n$. This will allow us to obtain more precise bounds on the dimension, since, once again, the contribution of oscillating terms will be (for a.e.~$k$) negligible in comparison with that of non-oscillating terms. This will be proved by the same methods as above. We will replace the oscillating term $\sin^{a}(2\thh(x_n,k))\sin^{2b}(\thh(x_n,k))$ by a more regular term $\sin^{a}(2\thh(x_n,k))\sin^{2b}(\thh(x_n,k))-Q_{n,a,b}(k)$, with $Q_{n,a,b}(k)$ small. The fact that we now have an infinite number of terms will not cause problems because the sum of their amplitudes converges absolutely.

First we need to do what we have already done once. We will split each non-oscillating term in two, a constant and an oscillating term. The latter will be treated as the other oscillating terms. We define
\[
F_{a,b}\equiv \frac 1\pi\int_0^{\pi} \sin^{a}(2\tht)\sin^{2b}(\tht)\,d\tht.
\]
If $2\nmid a$, then this is 0 and 
\[
W_{n,a,b}(k)\equiv \sin^{a}(2\thh(x_n,k))\sin^{2b}(\thh(x_n,k))
\]
is oscillating. If $2|a$, the corresponding term is non-oscillating  and we split it into $F_{a,b}$ and 
\[
W_{n,a,b}(k)\equiv \sin^{a}(2\thh(x_n,k))\sin^{2b}(\thh(x_n,k))-F_{a,b}.
\]
Let
\[
F(v_k)\equiv \sum_{\substack{a,b\ge 0 \\ a+b\ge 1}} \frac{(-1)^{a+b+1}}{2a+2b}
\binom{a+b}a v_k^{a+2b} F_{a,b}
\]
\[
G(v_k)\equiv \sum_{\substack{a,b\ge 0 \\ a+b\ge 1}} \frac{1}{2a+2b}
\binom{a+b}a |v_k|^{a+2b}.
\]
Then $F(v_k)$ is the sum of all the constant terms, and $G(v_k)$ is strictly larger than the sum of the amplitudes of the oscillating terms (for fixed $n$). Notice that 
\[
G(v_k)=-\tfrac 12\ln(1-|v_k|-v_k^2)
\]
and (see \cite{Pe,KR})
\[
F(v_k)=\frac 1\pi \int_0^\pi \frac 12\ln\Bigl(1+v_k\sin(2\tht) + v_k^2\sin^2(\tht)\Bigr)\,d\tht = \frac 12\ln\left(1+\frac{v_k^2}4\right).
\]
We take $\ga$ large so that $C_7G(v_k)/\pi<F(v_k)$ for all $k\in I$.

Now fix any $a,b$ and consider all the $W_{n,a,b}$'s. 
We can do everything as above. Take the intervals $I_{n,j}$ (these will be different for $2|a$ and $2\nmid a$, just as they were for the $\sin$ and $\cos$ terms). As before, the integral of $W_{n,a,b}$ over any $I_{n,j}$ is close to 0 (it would be exactly 0 if $\thh'(x_n,k)$ were constant on $I_{n,j}$). Thus from (\ref{3.2}$'$) we know that there is $Q_{n,a,b}(k)$ such that $|Q_{n,a,b}(k)|\le C_7/\pi$, and for $X_{n,a,b}(k)\equiv W_{n,a,b}(k)-Q_{n,a,b}(k)$ we have
\[
\int_{I_{n,j}}X_{n,a,b}(k)\,dk=0.
\]

As before, one can use this to prove that
\begin{equation} \lb{4.4}
\frac{\sum_{n=1}^{N} X_{n,a,b}}{N}\to 0
\end{equation}
for a.e.~$k$.
This holds for any $a,b$. Since the number of these pairs is countable, we have that for a.e.~$k$ \eqref{4.4} holds for any $a,b$. Using the fact that the sum of the amplitudes of the oscillating terms is finite, we obtain that 
\[
\frac 1N \sum_{n=1}^{N} \sum_{\substack{a,b\ge 0 \\ a+b\ge 1}}
\frac{(-1)^{a+b+1}}{2a+2b}\binom{a+b}a v_k^{a+2b}X_{n,a,b} \to 0
\]
for a.e.~$k$. Then for each such $k$ and for large $n>n_1(k)$ we have
\[
R_k(x_n+1)\in(e^{c_1 n},e^{c_2 n}),
\]
where $c_i= c_i(k)\equiv F(v_k)+(-1)^iC_7G(v_k)/\pi$ (remember that $G(v_k)$ is strictly larger than the sum of the amplitudes). We know from \cite{Si} that for a.e.~$\phi$ the spectral measure $\mu_{\phi}$ is supported on the set of these $k$'s because it is a set of full measure.

Let us now estimate $C_7$ for $|v_k|+v_k^2<1$. Let $I$ be a small interval around $k$, so that $w_2$ is arbitrarily close to $|v_k|$. Then $C_4-C_3<2(|v_k|+v_k^2)$ and $C_4<2$, and one can pick $D_1,D_2$ so that $(D_2-D_1)/D_1< 2(|v_k|+v_k^2)/(\ga-2)$ and $D_2/D_1<\ga/(\ga-2)$. We trivially have $\pi(2w_2+w_2^2)/(\ga^2-2)<7(|v_k|+v_k^2)/\ga^2$ for small $I$ and $\ga\ge 5$. So we obtain
\begin{equation} \lb{4.5}
C_7<\min\left\{\frac{28(|v_k|+v_k^2)}{(\ga-2)^2}, \frac{2(|v_k|+v_k^2)}{\ga-2}\right\}\equiv C_7'.
\end{equation}
Let $I$ be small enough so that for any $k'\in I$
\[
c_1(k'),c_2(k')\in \left(F(v_k)-\frac{C_7'}\pi G(v_k),F(v_k)+\frac{C_7'}\pi G(v_k)\right).
\]
Repeating the proof of {\it Theorem \ref{T4.1}\/} we obtain

\begin{theorem} \lb{T4.2}
With the notation of {\it Theorem \ref{T4.1}\/} let $\ga\ge 5$, $F(x)=\frac 12\ln(1+\frac {x^2}4)$ and $G(x)=-\frac 12\ln(1-|x|-x^2)$. For $k\in(0,\pi)$
let $E(k)=2\cos(k)$, $v_k=-\frac v{\sin(k)}$,
\[
\alpha_1(k)\equiv 1-2\, \frac{F(v_k)+\frac{28(|v_k|+v_k^2)}{\pi(\ga-2)^2}G(v_k)}{\ln(\ga)},
\]
\[
\alpha_2(k)\equiv 1-2\, \frac{F(v_k)-\frac{28(|v_k|+v_k^2)}{\pi(\ga-2)^2}G(v_k)}{\ln(\ga)}.
\]
Then for a.e.~$\phi$ we have for all $k$ such that $|v_k|+v_k^2<1$ and
$\alpha_2(k)<1$, that on a small interval around $E(k)$ the measure $\mu_{\phi}$ is $\alpha_1(k)$-continuous and $\alpha_2(k)$-singular.
\end{theorem}

\smallskip
\noindent{\it Remarks.}
1. Notice that the hypotheses imply that $\alpha_1(k)>0$ whenever $\alpha_2(k)<1$.
\smallskip

\noindent
2. So under the above conditions we know that for a.e.~$\phi$, the local dimension of the spectral measure is within an interval which is close to 1, and for large $\ga$ its size is small compared to its distance from 1.


\section{Fractional Dimension for all Boundary Conditions}  \lb{S5}

Now we turn to the proof for {\it all\/} boundary conditions. We return to considering the term $g_n(k)$ and three oscillating terms first. We again restrict our considerations to the $\sin$ term, the two $\cos$ terms are treated similarly. 

It turns out that the $X_n$ from Section \ref{S4} are not regular enough to prove \eqref{2.6} (with $\sin$ replaced by $X_n$). Therefore we need to make a new breakup of the oscillating term to obtain a more regular $X_n$. This is done in the Appendix in the proof of {\it Theorem \ref{TA.1}} (with $f_n(k)=2\thh(x_n,k)$, $\beta=2$, $\del_1=1-D_1$ and $\del_2=D_2-1$). Using the reasoning from Section \ref{S2} we then obtain our main result:

\begin{theorem} \lb{T5.1}
Let $H_{\phi}$ be the discrete Schr\" odinger operator on $\bbZ^+$ with potential $V_{v,\ga}$ given by \eqref{1.6}, and boundary condition $\phi$. Let $\mu_{\phi}$ be its spectral measure. For any closed interval of energies $J\subset(-2,2)$ there is $v_0>0$ and $\ga_0\in\bbN$ such that if $0<|v|<v_0$ and $\ga\ge\ga_0v^{-2}$ is an integer, then for any $\phi$, the measure $\mu_{\phi}$ has fractional Hausdorff dimension in $J$.
\end{theorem}

\smallskip
\noindent{\it Remark.}
A priori, $\ga_0$ depends on $v$. However, \eqref{5.1} below gives $\eps=O(v)$ as $v\to 0$ (since $C_1=O(v^2)$, $C_7=O(v\ga^{-2})$ and $w_2=O(v)$). So if $\ga=O(\eps^{-2})=O(v^{-2})$, then in \eqref{5.2} $\alpha(\eps)<1$ (because $D_1,D_2\to 1$ as $v\to 0$). Compare with the remark after {\it Theorem \ref{T4.1}}.
\smallskip

\begin{proof}
Let $I\subset(0,\pi)$ be such that $2\cos(I)=J$. Let $\alpha_1'\equiv 1-(w_2+w_2^2)/\ln(\ga)>0$ (with $w_2+w_2^2<1$ and $\ga\ge 3$). We already know that by {\it Proposition \ref{P1.2}\/} $\mu_{\phi}$ is $\alpha_1'$-continuous on $I$.

By the Appendix, the absolute value of the ``small'' term $\sin(2\thh(x_n,k))-X_n(k)$ from the new breakup is at most $\tfrac \pi 2(\del_1+\del_2)(1-\del_1)^{-1} \le \tfrac \pi 2 C_7$,
and for all $\eps> \eps_0=\eps_0(D_1,D_2,\ga)\equiv \bigl(2\ln(1+\tfrac{10}\ga D_2D_1^{-1})\bigr)^{1/2}$ there is $A^\eps$ with $\dim(A^\eps)<1$ such that \eqref{2.6} with $X_n(k)$ 
in place of $\sin(2\thh(x_n,k))$ is satisfied. Here $C_7,\eps_0\to 0$ as $D_1,D_2\to 1$ and $\ga\to \infty$, which is guaranteed when $\ga\to\infty$. Hence we proceed as follows.                                                                                                                  

This time we let $\ga$, $D_1$ and $D_2$ be such that
\[
\frac\pi 2 C_7\left(\frac{w_2}2+\frac{3w_2^2}8\right)<C_1
\]
(this is possible for all large $\ga$). Then for $n>n_0$ we have
\[
\left|\sum_{n=n_0+1}^N\frac{v_k}2\sin(2\thh(x_n,k))\right|\le\frac{w_2}2 \frac\pi 2 C_7 N + \left|\frac{v_k}2\sum_{n=n_0+1}^N X_n(k)\right|.
\]
We do the same with the other two oscillating terms, and the three terms containing $C_7$ add up to 
\[
\frac\pi 2 C_7\left(\frac{w_2}2+\frac{3w_2^2}8\right)N < C_1N.
\]

Next, we choose $\eps$ with
\begin{equation} \lb{5.1}
0<\eps\left(\frac{w_2}2+\frac{3w_2^2}8\right)<C_1-\frac\pi 2 C_7\left(\frac{w_2}2+\frac{3w_2^2}8\right)
\end{equation}
and we take $\ga$ large enough so that $\eps_0<\eps$. By the Appendix there is a set $A^\eps$ such that 
\begin{equation} \lb{5.2}
\dim(A^\eps)\le \alpha(\eps) \equiv 1-\frac{\eps^2-2\ln\left(1+\frac{10D_2}{\ga D_1}\right)}{2\ln(\ga)}
\end{equation}
and \eqref{2.6} holds with $\sin$ replaced by $X_n$. Similarly for the other two oscillating terms, and we let $A^\eps$ be the union of the three sets.

From the above discussion it follows that for $k\in I\backslash A^\eps$ and $n$ large enough we have
\begin{equation} \lb{5.3}
R_{k}(x_n+1)>e^{c_1 n}=x_n^{\beta_1},
\end{equation}
where 
\[
c_1=C_1-\left(\frac\pi 2 C_7+\eps\right)\left(\frac{w_2}2+\frac{3w_2^2}8\right) > 0
\]
and $\beta_1=c_1/\ln(\ga)$. It is known (see \cite{JL}) that for a.e.~$k$ w.r.t.~$\mu_0$ we have
\[
\lims_{L\to\infty} \frac{\|R_k\|_L}{L^{1/2}\ln(L)} <\infty.
\]
Given \eqref{5.3}, this can be true only if
$\mu_0(I\backslash A^\eps)=0$. Thus $\mu_0(I\cap\cdot)$ is supported on $A^\eps$.

Hence on $I$ the measure $\mu_{0}$ is $(\dim(A^\eps)+\delta)$-singular for any $\delta>0$. Since $D_1$, $D_2$ can be arbitrarily close to $C_5$, $C_6$, and since $C_6/C_5<\ga/(\ga-2)$ if $w_2+w_2^2<1$, we have
\[
\dim(A^\eps)< \alpha_2'\equiv 1-\frac{\eps^2-2\ln\left(1+\frac{10}{\ga-2}\right)}{2\ln(\ga)}.
\]
So $\mu_0$ is $\alpha_2'$-singular on $I$. If in this argument we replace $u_k$ by the generalized eigenfunction for the same energy satisfying boundary condition $\phi$, we obtain the same singularity for $\mu_{\phi}$.
Now take $\ga$ large so that $\alpha_2'<1$, and the result follows. 
\end{proof}

At this point we can do the same as what we did after proving {\it Theorem \ref{T4.1}}: consider an infinite sum of terms instead of $g_n(k)$. 

Now we use a different type of regularization of the oscillating terms, but everything can be done as before, with one adjustment. 
We need to use {\it Theorem \ref{TA.3}\/} in place of {\it Theorem \ref{TA.1}}, which gives us a different bound for the difference 
of the oscillating term $\sin^{a}(2\thh(x_n,k))\sin^{2b}(\thh(x_n,k))$ and its regularization $X_{n,a,b}(k)$, namely $\tfrac \pi 2 (a+b)C_7$.
The derivative of the oscillating term enters here, and we use the (very crude) estimate 
\[
\big\|[\sin^a(2\tht)\sin^{2b}(\tht)]'\big\|_\infty \le (a+b) \big\|[\sin(2\tht)]'\big\|_\infty.
\]
                                                                              
So this time we need to take $F(v_k)$ and $G(v_k)$ as before (the latter will be the coefficient for $\eps$), and also
\[
\tilde G(v_k)=\sum_{\substack{a,b\ge 0 \\ a+b\ge 1}} \frac{1}{2} \binom{a+b}a |v_k|^{a+2b}
=\frac 12 \,\,\frac{|v_k|+v_k^2}{1-|v_k|-v_k^2}
\]
which will be the coefficient for $\frac\pi 2 C_7$. We take $\ga$ large so that $\frac\pi 2 C_7\tilde G(v_k)<F(v_k)$,
and take $\eps>0$ such that 
\[
\eps G(v_k)<F(v_k)-\frac\pi 2 C_7 \tilde G(v_k).
\]
Then given any pair $a,b$ we construct the set $A_{a,b}^\eps$ for $X_{n,a,b}$. 
We show, as above, that its dimension is at most $\alpha(\eps)$. Letting the (countable) union of 
these sets play the role of $A^\eps$ in the proof of {\it Theorem \ref{T5.1}\/}, and considering \eqref{4.5}, we obtain

\begin{theorem} \lb{T5.2}
With the notation of {\it Theorem \ref{T5.1}\/} let $\ga\ge 5$. For $k\in(0,\pi)$
let $E(k)=2\cos(k)$, $v_k=-\frac v{\sin(k)}$,
\[
\eps_k\equiv \frac{\ln\left(1+\frac{v_k^2}4\right)-\frac{14\pi(|v_k|+v_k^2)^2}{(\ga-2)^2(1-|v_k|-v_k^2)}}{-\ln(1-|v_k|-v_k^2)},
\]
\[
\alpha_1'(k)\equiv 1-\frac{|v_k|+v_k^2}{\ln(\ga)},
\]
\[
\alpha_2'(k)\equiv 1-\frac{\eps_k^2-2\ln\left(1+\frac{10}{\ga-2}\right)}{2\ln(\ga)}.
\]
If $k$ is such that $|v_k|+v_k^2<1$, $\eps_k>0$ and $\alpha_2'(k)<1$, then
on a small interval around $E(k)$ each $\mu_{\phi}$ is $\alpha_1'(k)$-continuous and $\alpha_2'(k)$-singular.
\end{theorem}

\smallskip
\noindent{\it Remarks.}
1. Notice that the hypotheses imply that $\alpha_1'(k)>0$.
\smallskip

\noindent
2. So for large $\ga$, the dimension of each spectral measure $\mu_{\phi}$ is within an interval which is close to $1$, and its size is comparable to its distance from 1. The estimate for a.e.~$\phi$ in Section~\ref{S4} is better by a factor of $\ga^{-2}$.
\smallskip

To illustrate the obtained results we provide an

\begin{example}
Let us assume that $v$, $\ga$ and $I$ are such that $\ga\ge 10^4v^{-2}$, and $|v_k|\le \frac 12$ for any $k\in I$. Estimating the quantities in {\it Theorems \ref{T4.2} \rm and \it \ref{T5.2}}, one can obtain
\[
[\alpha_1(k),\alpha_2(k)]\subseteq 
\left[1-\frac{\ln\left(1+\frac{v_k^2}{4}\right)}{\ln(\ga)}-\frac{10}{\ga^2\ln(\ga)}, \,
1-\frac{\ln\left(1+\frac{v_k^2}{4}\right)}{\ln(\ga)}+\frac{10}{\ga^2\ln(\ga)}\right]
\]
and
\[
[\alpha_1'(k),\alpha_2'(k)]\subseteq \left[1-\frac {|v_k|+v_k^2}{\ln(\ga)}, \, 1-\frac{v_k^2}{500\ln(\ga)}\right]
\]
(notice that $v_k=-2v/\sqrt{4-E^2}$).
For instance, take $v=\frac 1{10}$, $\ga=10^6$ and $J=[-1.9,1.9]$. Then for a.e.~$\phi$ the local dimension of $\mu_{\phi}$ at any $E\in J$ is in
\[
\left[1-\frac{\ln\left(1+\frac{1}{100(4-E^2)}\right)}{6\ln(10)}-\frac{1}{10^{12}} ,\,
1-\frac{\ln\left(1+\frac{1}{100(4-E^2)}\right)}{6\ln(10)}+\frac{1}{10^{12}}\right]
\]
and for all $\phi$ it is in
\[
\left[1-\frac 1{20\ln(10)\sqrt{4-E^2}} ,\, 1-\frac 1{10^5\ln(10)(4-E^2)}\right]\subseteq
\left[1-\frac 1{10} ,\, 1-\frac 1{10^6}\right].
\]
\end{example}



\section{Random Operators} \lb{S6}

In this section we consider potentials of the form \eqref{1.6}, with certain randomness in the {\it position\/} of the special sites $x_n$. More precisely, $x_n$ will be a random variable uniformly distributed over $\{\ga^n-n,\dots,\ga^n+n\}$. The fact that the size of these sets grows will yield certain ``averaging'' of $\ln(R_k(x_n+1))-\ln(R_k(x_{n-1}+1))$ and thus a constant growth (in the limit $n\to\infty$) of $\ln(R_k(x_n+1))$ for a.e.~realization of the potential. As a consequence we will be able to compute the exact dimension of the spectral measures for these random potentials. 

We begin with a standard result.

\begin{lemma} \lb{L6.1}
For a.e.~$x\in\bbR$ there is $q_0$ such that for any integer $q>q_0$ we have
\begin{equation} \lb{6.1}
\dist(qx,\bbZ)>\frac 1{q^2}.
\end{equation}
\end{lemma}

\begin{proof}
For any $n\in\bbN$ the measure of the set of $x\in[n,n+1)$, for which \eqref{6.1} fails, is $2q^{-2}$. This is summable in $q$, and so by the Borel-Cantelli Lemma the measure of those $x\in[n,n+1)$ for which \eqref{6.1} fails infinitely often is zero.
\end{proof}

Let $\calX$ be the set of all such $x$ and let $\calK=2\pi\calX$. Notice that $\calK$ is a set of full measure, not intersecting $2\pi\bbQ$. The ``averaging'' result we need is also well-known.

\begin{lemma} \lb{L6.2}
Let $k\in\calK$ and let $f\in C^4(\bbR)$ have period $2\pi$ with $\int_0^{2\pi}f(t)\,dt=0$. Then there is $C=C(k,f)<\infty$ such that for every $\tht\in\bbR$ and $n\in\bbN$
\[
\left|\sum_{\ell=1}^n f(\tht+\ell k)\right|\le C.
\]
\end{lemma}

\begin{proof}
Let $\sum_{q=-\infty}^\infty a_q e^{iqt}$ be the Fourier series of $f(t)$.
Since $f$ is $C^4$, we know that $|a_q|\le cq^{-4}$ for some $c$. Thanks to
this fact all the sums appearing in this argument are pointwise absolutely
convergent. Also, $a_0=0$ by the hypothesis.
We have
\begin{align*}
\allowdisplaybreaks
\left|\sum_{\ell=1}^n f(\tht+\ell k)\right| \le \sum_q |a_q| \left|\sum_{\ell=1}^n e^{iq\ell k}\right|
& = \sum_{q\neq 0} |a_q|\left|\frac{1-e^{iqnk}}{1-e^{iqk}}\right|
\\ & \le \pi \sum_{q\neq 0} \frac{|a_q|}{\dist(qk,2\pi\bbZ)}
\le c \sum_{q=1}^\infty \frac{1}{q^4\dist(qx,\bbZ)}
\end{align*}
where $x=k/2\pi\in\calX$. By {\it Lemma \ref{L6.1}\/} this is bounded by
\[
c'+c \sum_{q=1}^\infty \frac1{q^2} \equiv C < \infty
\]
where $c'<\infty$ depends on $x$ and $c$.
\end{proof}

Let $\omega_n$ be a random variable uniformly distributed over $\{-n,-n+1,\dots,n\}$ and let $(\Omega,\nu)$ be the product probability space for these $\omega_n$'s ($n\in\bbN$). For $\omega=(\omega_1,\omega_2,\dots)\in\Omega$ let
\begin{equation} \lb{6.2}
V_{v,\ga}^{(\omega)}(x)\equiv \begin{cases} v & \text{$x=x_n^{(\omega)}\equiv \ga^n+\omega_n$ for some $n\ge 1$},
\\ 0 & \text{otherwise}. \end{cases}
\end{equation}

It is a consequence of the ``smallness'' of $\omega_n$ (compared to $\ga^n$), that all previous results for $V_{v,\ga}$ apply to each $V_{v,\ga}^{(\omega)}$ as well. In this random case, however, we will prove and compute {\it exact\/} ({\it local\/}) {\it dimension\/} for the spectral measures for a.e.~$\omega$ and a.e.~boundary condition $\phi$. Moreover, our results will apply to any $v$, $k$ and $\ga$, not only to small $v_k$ and large $\ga$.

Let us fix $k\in\calK$ and define
\[
X_n(\omega)\equiv \tfrac 12\ln\bigg(1+v_k\sin\Bigl(2\thh_k\left(x_n^{(\omega)}\right)\Bigr) + v_k^2\sin^2\Bigl(\thh_k\left(x_n^{(\omega)}\right)\Bigr)\bigg)- \tfrac 12\ln\left(1+\frac{v_k^2}4\right).
\]
Obviously $|X_n(\omega)|<M$ for some $M=M(k)<\infty$.
We will prove that for a.e.~$\omega$
\begin{equation} \lb{6.3}
\frac{\sum_{n=1}^{N} X_n(\omega)}{N}\to 0,
\end{equation}
which in turn implies (see below) that the exact dimension of $\mu_{\phi}^{(\omega)}$ for a.e.~$\omega$ and a.e.~$\phi$ is
\[
1-\frac{\ln\left(1+\frac{v_k^2}4\right)}{\ln(\ga)}
=1-\frac{\ln\left(1+\frac{v^2}{4-E^2}\right)}{\ln(\ga)}.
\]

We use the same methods as earlier. We exploit {\it Lemmas \ref{L6.1}\/ \rm and \it \ref{L6.2}\/} to show that the expectations of the cross terms $X_mX_n$ for $m<n$ are small, which implies \eqref{6.3} for a.e.~$\omega$. This time, however, we consider expectations w.r.t.~$\omega$, rather than $k$.

Let
\[
f(\tht)\equiv \tfrac 12\ln\Bigl(1+v_k\sin(2\tht) + v_k^2\sin^2(\tht)\Bigr)- \tfrac 12\ln\left(1+\frac{v_k^2}4\right).
\]
Then $f(\tht)=f(\tht+2\pi)$, $\int_0^{2\pi}f(\tht)\,d\tht=0$, and $f$ is $C^4$, so $f$ is as in {\it Lemma \ref{L6.2}}. Let $\delta,\omega\in\Omega$ be such that $\delta_j=\omega_j$ for $j=1,\dots,n-1$, and $\delta_n=\omega_n+l$ for some $l$. Then the recursive relation for the EFGP transform of generalized eigenfunctions implies that if $X_n(\omega)=f(\tht)$ for some $\tht$, then $X_n(\delta)=f(\tht+lk)$. This shows why {\it Lemma \ref{L6.2}\/} enters in our argument.

For $m<n$ we have with $\bbE = \bbE_\omega$
\begin{align*}
\allowdisplaybreaks
|\bbE(X_mX_n)| &\le \max_{c_1,c_2,\dots,c_{n-1}}\Big\{ \, \Big|\bbE\bigl(X_mX_n \,|\, \omega_1=c_1,\dots,\omega_{n-1}=c_{n-1}\bigr)\Big| \, \Big\}
\\ &\le \max_{c_1,c_2,\dots,c_{n-1}}\Big\{ M \Big|\bbE\bigl(X_n \,|\, \omega_1=c_1,\dots,\omega_{n-1}=c_{n-1}\bigr)\Big| \, \Big\}.
\end{align*}
By {\it Lemma \ref{L6.2}\/} there exists $C<\infty$ such that the last expectation is at most $C/(2n+1)$. Hence for fixed $k\in\calK$ there is $D=D(k)<\infty$ such that for $m<n$
\[
|\bbE(X_mX_n)|<\frac Dn.
\]
Then
\[
\bbE\left(\left|\sum_{n=1}^N X_n\right|\right)\le \bbE\left(\left(\sum_{n=1}^N X_n\right)^2\right)^{1/2}\le (M^2N+2DN)^{1/2}=cN^{1/2}.
\]
As before, using summability of $\bbE\left(\left|N^{-4} \sum_{n=1}^{N^4} X_n\right|\right)$ and boundedness of $X_n$, one shows that \eqref{6.3} holds for a.e.~$\omega$.

We have proved \eqref{6.3} for a.e.~$k$ and a.e.~$\omega$. By the Fubini Theorem, \eqref{6.3} holds for a.e.~$\omega$ and a.e.~$k$. Thus for these $(\omega,k)$ we obtain an exact power of the growth for the solution $u_k$. Namely we have that for any 
\[
c_1<\frac 12\ln\left(1+\frac{v_k^2}4\right)<c_2
\]
there is $n_0$ such that for $n>n_0$
\[
R_k\left(x_n^{(\omega)}+1\right)\in(e^{c_1 n},e^{c_2 n}).
\]
Also, {\it Lemma \ref{L2.1}\/} gives us the existence of a subordinate
solution  $u_k^{\text{sub}}\sim(P_k,\psi_k)$ for energy $2\cos(k)$, so that for large $n$
\[
P_k\left(x_n^{(\omega)}+1\right)\le e^{-c_1n}.
\]

Since for any such $\omega$, for a.e.~$\phi$ the spectral measure $\mu_{\phi}^{(\omega)}$ is supported on the set of the corresponding $k$'s, one can conclude (by methods presented earlier) the following

\begin{theorem} \lb{T6.3}
Let $(\Omega,\nu)$ be the above probability space. For $\omega\in\Omega$ let $H_{\phi}^{(\omega)}$ be the discrete Schr\" odinger operator on $\bbZ^+$ with potential $V_{v,\ga}^{(\omega)}$, given by \eqref{6.2} with $v\neq 0$ and $\ga>1$, and boundary condition $\phi$. Let $\mu_{\phi}^{(\omega)}$ be its spectral measure. Let
\[
J\equiv \left( -\sqrt{4-\frac{v^2}{\ga-1}},\,\sqrt{4-\frac{v^2}{\ga-1}}\,\right)
\]
if $v^2<4(\ga-1)$, and $J\equiv \emptyset$ otherwise.
Then for a.e.~$\omega$ and for a.e.~$\phi$ the measure $\mu_{\phi}^{(\omega)}$ is purely singular continuous in $J$, with local dimension
\[
1-\frac{\ln\left(1+\frac{v^2}{4-E^2}\right)}{\ln(\ga)},
\]
and it is dense pure point in the rest of the interval $[-2,2]$. 
\end{theorem}


\section{Whole Line Operators With Symmetric Potentials}  \lb{S7}

It is readily seen that our results also apply to certain whole-line operators (satisfying \eqref{1.1} for $x\in\bbZ$). Let us consider the operator $\tilde H$ on $\ell^2(\bbZ)$ with potential $\tilde V_{v,\ga}$ given by \eqref{1.6} for $x\ge 1$ and by $\tilde V_{v,\ga}(x)=\tilde V_{v,\ga}(1-x)$ for $x\le 0$. This potential is reflected about $\frac 12$. One easily sees that $\tilde H=\tilde H\res_\calE\oplus \tilde H\res_\calO$ where $\calE=\{u\in \ell^2(\bbZ)\,\mid\, u(1-x)=u(x) \text{ for all } x\in\bbZ\}$ and $\calO=\{u\in \ell^2(\bbZ)\,\mid\, u(1-x)=-u(x) \text{ for all } x\in\bbZ\}$ are, respectively, the even and odd subspaces of $\ell^2(\bbZ)$.

On the other hand, if $\delta_1$ is the delta function at $x=1$, then 
\begin{equation} \lb{7.1}
\begin{aligned}
\tilde H\res_\calE &\cong H_0+\delta_1=H_{\frac{3\pi}4},
\\ \tilde H\res_\calO &\cong H_0-\delta_1=H_{\frac\pi 4}
\end{aligned}
\end{equation}
by simply taking the restriction of $u$ to $\bbZ^+$. Hence $\tilde \mu=\mu_{\pi/4}+\mu_{3\pi/4}$ is a spectral measure for $\tilde H$, and all we have proved about the measures for the half-line operators applies to $\tilde \mu$. Notice that variation of boundary condition (i.e., of $V(1)$) in the half-line case correponds to varying $\tilde V(1)=\tilde V(0)$, as can be seen from \eqref{7.1}. We conclude:

\begin{theorem} \lb{T7.1}
Let $\tilde H$ be the discrete Schr\" odinger operator on $\bbZ$, with potential $\tilde V_{v,\ga}$ given by \eqref{1.6} for $x\ge 1$ and reflected about $\frac 12$. Let $\tilde H_{\phi}\equiv \tilde H-\tan(\phi)(\delta_0+\delta_1)$ and let $\tilde \mu_{\phi}$ be the spectral measure of $\tilde H_{\phi}$. Then the statements of {\it Theorems \ref{T4.1}, \ref{T4.2}, \ref{T5.1}\/} and {\it \ref{T5.2}\/} hold for $\tilde H_{\phi}$.

If instead of $\tilde H$ and $\tilde V_{v,\ga}$ we consider $\tilde H^{(\omega)}$ and $\tilde V_{v,\ga}^{(\omega)}$ {\rm (}given by \eqref{6.2} and reflected about $\frac 12${\rm )}, and $\tilde H_{\phi}^{(\omega)}\equiv \tilde H^{(\omega)}-\tan(\phi)(\delta_0+\delta_1)$, then the statement of {\it Theorem \ref{T6.3}\/} holds for $\tilde H_{\phi}^{(\omega)}$.
\end{theorem}

\medskip
\appendix
\section*{Appendix. A Dimensional Estimate} \lb{SA}
\renewcommand{\theequation}{A.\arabic{equation}}
\renewcommand{\thetheorem}{A.\arabic{theorem}}
\setcounter{theorem}{0}
\setcounter{equation}{0}

\begin{theorem} \lb{TA.1}
Let $\ga>1$, $\beta>0$ and $\del_1,\del_2\in [0,1)$. Let $I$ be a finite 
interval and let $f_n\in C(I)$ be such that 
\begin{equation} \lb{A.1}
\frac{f_n'(k)}{\beta\ga^n} \in [1-\del_1,1+\del_2]
\end{equation}
for all $n\in\bbN$ and a.e.~$k\in I$. Let
\[
F(k)\equiv 
\lims_{N\to\infty}\frac 1N \left|\sum_{n=1}^N \sin\bigl(f_n(k)\bigr) \right| 
\]
and $A^\eps\equiv\{k \,|\, F(k)\ge\eps\}$. Then there is
$\eps_0= \eps_0(\del_1,\del_2,\ga)$ with $\eps_0\to 0$ as 
$\del_1,\del_2,\ga^{-1}\to 0$ such that for $\eps>\eps_0$ the set $A^\eps$ has 
Hausdorff dimension less than $1$. 
\end{theorem}

\noindent
{\it Remarks.} 
1. In fact, we obtain
\[ 
\dim(A^\eps)
\le 1- \frac{ \left(\eps-\frac \pi 2 \frac{\del_1+\del_2}{1-\del_1} \right)^2 
- 2\ln \left( 1+ \frac{10}\ga \frac{1+\del_2}{1-\del_1} \right) } {2\ln(\ga)} 
\] 
whenever $\eps>\tfrac \pi 2 (\del_1+\del_2)(1-\del_1)^{-1}$.
\smallskip

\noindent
2. Similar questions have been studied using dynamical systems and Riesz measures (see, e.g., \cite{FS,Per}). The methods, however, seem to require $\delta_1=\delta_2=0$ and $\ga\in\bbN$.
\smallskip

\noindent
3. This result is only useful if we can take $\eps$ smaller than 1. Hence even in the case $\del_1=\del_2=0$ it applies only when $\ga$ is large (larger than $10/(\sqrt e -1)$). 
\smallskip

\noindent
4. See Section \ref{S5} for the application of this result in the present paper.
\smallskip

The rest of this appendix is devoted to the proof of the above estimate.
First, we explain the presence of the term $\tfrac \pi 2
(\del_1+\del_2)(1-\del_1)^{-1}$. The reason is the same as in Section \ref{S4}.
We need more regularity than the function $\sin(f_n(k))$ possesses, and so we
will need to break it up in two terms. One of them will be small, with absolute
value not more than $\tfrac \pi 2 (\del_1+\del_2)(1-\del_1)^{-1}$, whereas the
other one will be ``regular'' enough so that we will be able to prove for it the
above theorem with $(\eps-\dots)^2$ replaced by just $\eps^2$. These two facts
then yield the theorem as stated. We note that the regular term, denoted $X_n$, will 
be different from the $X_n$ term from Section \ref{S4}, which is not regular
enough for the purposes of this argument.

Before we perform this breakup, notice that if we define intervals $I_{n,j}=[k_{n,j},k_{n,j+1}]$ in
the same way as in Section \ref{S3}, but with $2\thh(x_n,k)$ replaced by $f_n(k)$ (i.e., so that $f_n(k_{n,j})$ are
multiples of $2\pi$), then \eqref{A.1} implies
\begin{equation} \lb{A.2}
|I_{n,j}|\in \left[ \frac{2\pi} {\beta D_2\ga^{n}}, \frac{2\pi} {\beta 
D_1\ga^{n}} \right]
\end{equation}
with $D_1\equiv 1-\del_1$ and $D_2\equiv 1+\del_2$.

We now define for $k\in I$
\[
\varphi_n(k)\equiv
\begin{cases} f_n(k_{n,j})+2\pi\frac{k-k_{n,j}}{|I_{n,j}|} & k\in
I_{n,j}
\\ f_n(k) & k\in I\backslash [K_{n,1},K_{n,2}].
\end{cases}
\]
Notice that $\varphi_n(k_{n,j})=f_n(k_{n,j})$ for all $j$, and
$\varphi_n(k)$ is linear on each $I_{n,j}$. So if we let $X_n(k)\equiv
\sin(\varphi_n(k))$, then $X_n$ is a series of exact sin waves on intervals 
$I_{n,j}$. This is the type of regularity
we need and $X_n$ will be the regular term in our breakup.

Now we want to estimate
the small term $\sin(f_n(k))-X_n(k)$. To do that, we need
an upper bound on $|f_n(k)-\varphi_n(k)|$. This will be maximal if $f_n'(k)$
equals $\beta D_1\ga^n$ on some interval $(k_{n,j},k_{n,j}+c)$ and
$\beta D_2\ga^n$ on $(k_{n,j}+c,k_{n,j+1})$ (or vice-versa), and the maximum
will occur at $k_{n,j}+c$. Since in such case
$\beta D_1\ga^nc+\beta D_2\ga^n(|I_{n,j}|-c)=2\pi$ by the
definition of $k_{n,j}$, we can compute $|I_{n,j}|$ and $\varphi_n$ in terms of
$c$, and then maximize for $c$. We obtain
\[
|f_n(k)-\varphi_n(k)|\le
2\pi \frac{\sqrt{D_2}-\sqrt{D_1}} {\sqrt{D_2}+\sqrt{D_1}}
\]
which yields the above claimed estimate
\[
|\sin(f_n(k))-X_n(k)| \le |f_n(k)-\varphi_n(k)| \le 2\pi \frac{D_2-D_1}{4D_1}
\le \frac \pi 2 \frac{\del_1+\del_2}{1-\del_1}.
\]

Now we only need to treat the term $X_n$. We start with a technical

\begin{lemma} \lb{LA.2}
There is a constant $c_0$ such that for any $n\ge 1$ and any $0\le\eps\le \frac 12$ we have
\[
\sum_{j=0}^{\lfloor\frac n2 - \eps n\rfloor} \binom{n}{j}\le c_0n2^ne^{-2\eps^2n}.
\]
\end{lemma}

\noindent{\it Remark.}
By the normal approximation to the binomial distribution, the left-hand side is roughly $2^n\Phi(-2\eps\sqrt n)< 2^ne^{-2\eps^2n}$, where $\Phi$ is the standard normal distribution function. The extra factor $n$ is added for convenience of proof and can be removed.
\smallskip

\begin{proof}
The sum is obviously smaller than $n\binom n{\lfloor\frac n2 - \eps n\rfloor}$, so we will estimate this. By Stirling's formula we have for $n\ge 1$
\[
c_1 \sqrt n\left(\frac ne\right)^n<n!<c_2 \sqrt n \left(\frac ne\right)^n
\]
for some $c_1,c_2>0$. Let $\lfloor \frac n2-\eps n\rfloor = \big(\frac 12-\delta\big)n$ (hence $\eps\le\delta\le \frac 12$). Then
\[
\binom n {\big( \frac 12-\delta \big) n} 
\le \frac {c_2 \sqrt{n} \left( \frac ne \right)^n} {c_1 \left[ \frac {(\frac 12-\delta)n}{e} \right]^{(\frac 12-\delta)n} 
c_1 \sqrt{\frac n2} \left[ \frac {(\frac 12+\delta)n}{e} \right]^{(\frac 12+\delta)n}}
= c_0 \left[ \left( \frac 12-\delta \right)^{\frac 12-\delta} \left( \frac 12+\delta \right)^{\frac 12+\delta} \right]^{-n}
\]

Since $\eps\le\delta$, it is sufficient to prove that
\[
\left( \frac 12-\delta \right)^{\frac 12-\delta} \left( \frac 12+\delta \right)^{\frac 12+\delta} 
\ge 2^{-1}e^{2\delta^2}
\]
for $0\le\delta\le \frac 12$.
This can be done by taking the logarithm of both sides of the inequality and observing that the resulting quantities have the same value and first derivative for $\delta=0$, and the left hand side has a larger second derivative in $(0,\frac 12)$.
\end{proof}

Let us now study a simple example which will illustrate our strategy. Let us 
take $\tilde X_n(k)\equiv \sgn(\sin(2^n\pi k))$ for $k\in [0,1]$ with 
$\sgn(0)\equiv 1$. Thus ${\tilde X}_n$ takes only values $1$ and $-1$, 
alternatively on intervals of lengths $2^{-n}$. Let us denote 
\[ 
A_{\tilde X}^\eps\equiv 
\left\{k\Big|\lims_{N\to\infty} \frac {\sum_{n=1}^N {\tilde X}_n(k)}N> \eps\right\}. 
\] 
We will show that the dimension of $A_{\tilde X}^\eps$ is 
smaller than 1.

Let $S_N$ be the union of those intervals $[j2^{-N},(j+1)2^{-N}]$ (for
$j=0,\dots,2^N-1$), in which $\sum_{n=1}^N {\tilde X}_n(k)\ge \eps N$. Their number
equals the number of sequences of $N$ symbols from the alphabet $\{-1,1\}$ such
that the number of occurences of $-1$ is at most $\lfloor (1-\eps)\frac N2\rfloor$.
For any $N_1$ the set $\bigcup_{N_1}^\infty S_N$ is obviously a $(2^{-N_1})$-cover
of $A_{\tilde X}^\eps$, and we have by {\it Lemma \ref{LA.2}\/}
\begin{align*}
\allowdisplaybreaks
h^\alpha(A_{\tilde X}^\eps) 
& \le \lim_{N_1\to\infty}\sum_{N=N_1}^\infty 2^{-\alpha N}\sum_{n=0}^{\lfloor \frac{(1-\eps)}2 N\rfloor} \binom Nn 
\\ & \le \lim_{N_1\to\infty}\sum_{N=N_1}^\infty 2^{-\alpha N} c_0N2^Ne^{-\frac {\eps^2}2 N}
\\ & = \lim_{N_1\to\infty}\sum_{N=N_1}^\infty c_0N \left(2^{1-\alpha}e^{-\frac{\eps^2}2}\right)^N. 
\end{align*} 
If $\alpha<1$ is such that $2^{1-\alpha}e^{-\frac{\eps^2}2}<1$, we get 
$h^\alpha(A_{\tilde X}^\eps)=0$.

Since we can do the same for
\[
B_{\tilde X}^\eps\equiv \left\{k\Big|\limi_{N\to\infty} \frac {\sum_{n=1}^N {\tilde X}_n(k)}N<
-\eps\right\}, \]
it follows, that the set
\[
\left\{k\Big|\lims_{N\to\infty} \left|\frac {\sum_{n=1}^N {\tilde X}_n(k)}N\right|> \eps\right\} 
= A_{\tilde X}^\eps\cup B_{\tilde X}^\eps
\]
has dimension at most $\alpha<1$.

We would like to prove now a similar result for our $X_n$'s. There is, however,
a problem. The technique used in the above
example was applicable to finite-valued functions only. Thus we have to
``discretize'' $X_n$ via another breakup into a ``small'' and a ``nice'' term.
Pick $p\in\bbN$ and define
\begin{equation} \lb{A.3}
Y_n(k)\equiv \frac{\lfloor pX_n(k)+\frac 12\rfloor}{p}.
\end{equation}
Then $Y_n$ takes values $\frac jp$ for $j=-p,\dots,p$ and $|X_n(k)-Y_n(k)|\le
\tfrac 1{2p}$. Later we will take $p\to\infty$, and then results which we prove
for $Y_n$ will apply to $X_n$ as well.

Finally we will break up $Y_n$ into $p$ even simpler terms. Let
\[
Y_{n,i}(k)\equiv \begin{cases} \sgn(Y_n(k)) & \text{if $|Y_n(k)|\ge\frac i{p}$,} \\ 0 & \text{otherwise,}\end{cases}
\]
for $i=1,\dots,p$. Then $Y_{n,i}$ takes only values $-1$, $0$ and $1$, and  
$Y_n(k)=\tfrac 1p\sum_{i=1}^{p}Y_{n,i}$.

It is obvious from the construction of $Y_{n,i}$ and the fact that $X_n$ is a perfect sin wave on $I_{n,j}$, that on any $I_{n,j}$ we have
\[
Y_{n,i}(k)=\begin{cases} 0 & k\in I_{n,j}^1(i)\equiv [k_{n,j},k_{n,j}+a_i|I_{n,j}|),
\\ 1 & k\in I_{n,j}^2(i)\equiv [k_{n,j}+a_i|I_{n,j}|,k_{n,j}+(\frac 12-a_i)|I_{n,j}|],
\\ 0 & k\in I_{n,j}^3(i)\equiv (k_{n,j}+(\frac 12-a_i)|I_{n,j}|,k_{n,j}+(\frac 12+a_i)|I_{n,j}|],
\\ -1 & k\in I_{n,j}^4(i)\equiv (k_{n,j}+(\frac 12+a_i)|I_{n,j}|,k_{n,j}+(1-a_i)|I_{n,j}|),
\\ 0 & k\in I_{n,j}^5(i)\equiv [k_{n,j}+(1-a_i)|I_{n,j}|,k_{n,j+1}], \end{cases}
\]
where $a_i=\tfrac 1{2\pi}\arcsin\bigl((i-\frac 12)/p\bigr)$.
Hence we have
\begin{equation} \lb{A.4}
\begin{gathered}
\frac{|I_{n,j}^1(i)\cup I_{n,j}^3(i)\cup I_{n,j}^5(i)|}{|I_{n,j}|}=4a_i,
\\ \frac{|I_{n,j}^2(i)|}{|I_{n,j}|}=\frac{|I_{n,j}^4(i)|}{|I_{n,j}|}=\frac 12-2a_i.
\end{gathered}
\end{equation}

To complete our argument, we will prove that for $\eps^2> 
2\ln(1+\tfrac {10}\ga D_2D_1^{-1})$ the sets 
\[
A_p^\eps\equiv \left\{k\Big|\lims_{N\to\infty} \frac {\sum_{n=1}^N Y_n(k)}N> \eps\right\},
\]
\[
B_p^\eps\equiv \left\{k\Big|\limi_{N\to\infty} \frac {\sum_{n=1}^N Y_n(k)}N< -\eps\right\}
\]
have dimension smaller than 1, and then deduce the same for
$X_n$. To this end it is sufficient to show that for any
$i=1,\dots,p$ the set 
\[
A_p^{\eps}(i)\equiv \left\{k\Big|\lims_{N\to\infty} \frac {\sum_{n=1}^N
Y_{n,i}(k)}N> \eps\right\} 
\] 
has dimension smaller than 1. This is because
\[
A_p^\eps\subseteq\bigcup_{i=1}^{p} A_p^{\eps}(i),
\]
and similarly for $B_p^\eps$.

Fix some $i$ and $\eps$. Let 
$I'$ be any closed sub-interval of $I$, not containing the endpoints of $I$. Let 
$N_0$ be such that for all $N> N_0$ we have $I'\subseteq(K_{N,1}, 
K_{N,2})$. Let us consider $\sum_{n=N_0+1}^{N+N_0} Y_{n,i}$ on $I'$ and denote 
\[
A\equiv \left\{k\in I'\Big|\lims_{N\to\infty} \frac {\sum_{n=N_0+1}^{N+N_0} Y_{n,i}(k)}N> \eps\right\}.
\]
Obviously $A=A_p^\eps(i)\cap I'$. Hence proving that $\dim(A)\le\alpha$ with 
$\alpha<1$ independent of $I'$ and $i$ will be enough to show 
$\dim(A_p^\eps)\le\alpha<1$.

As in our simple example, we will cover $A$ by a (recursively constructed) set of intervals.
Let $N\ge 1$ and define $A_{N+N_0}\equiv \{k\in I'|\sum_{n=N_0+1}^{N+N_0} Y_{n,i}(k)>\eps N\}$. Then for any $N_1\ge 1$ we have $A\subseteq\bigcup_{N=N_1}^\infty A_{N+N_0}$. If $k\in A_{N+N_0}$ for some $N$, then the sequence $\{s_n\}_{N_0+1}^{N+N_0}$ with $s_n\equiv Y_{n,i}(k)$ can have at most $\lfloor(1-\eps)N\rfloor$ zeros, and if it has $l$ zeros, then it can have at most 
\[
\left\lfloor \frac{N-l}2-\frac{\eps N}2\right\rfloor=\left\lfloor \frac{N-l}2-\frac{\eps N}{2(N-l)}(N-l)\right\rfloor
\]
occurences of $-1$. The number of such sequences (with $l$ zeros) is  at most
\[
\binom{N}l \sum_{n=0}^{\lfloor \frac{N-l}2-\frac{\eps N}{2(N-l)}(N-l) \rfloor} \binom {N-l}n 
\le \binom{N}l c_0(N-l) 2^{N-l}e^{-\frac{\eps^2N^2}{2(N-l)}}
\le c_0Ne^{-\frac{\eps^2}2 N}\binom Nl 2^{N-l}
\]
by {\it Lemma \ref{LA.2}}.

Let us pick one such sequence $\{s_n\}_{N_0+1}^{N+N_0}$ with $l$ zeros. We will construct a covering of the set of those $k\in I'$ which generate this sequence, by intervals $I_{N+N_0+1,j}$. Let $S_{N_0}$ be the union of those $I_{N_0+1,j}$ which have nonzero intersection with $I'$. Now construct inductively $S_n$ from $S_{n-1}$, so that $S_n$ is the union of those intervals $I_{n+1,j}$, which have nonzero intersection with the set
\[
\tilde S_n\equiv \begin{cases} 
\bigcup_{I_{n,j}\subseteq S_{n-1}} I_{n,j}^2(i) & \text{if } s_n=1,\\
\bigcup_{I_{n,j}\subseteq S_{n-1}} \bigl[I_{n,j}^1(i)\cup I_{n,j}^3(i)\cup I_{n,j}^5(i)\bigr] & \text{if } s_n=0,\\
\bigcup_{I_{n,j}\subseteq S_{n-1}} I_{n,j}^4(i) & \text{if } s_n=-1. \end{cases}
\]

We want to estimate $|S_{N+N_0}|$. This can be done recursively using \eqref{A.4} and these facts: 
\begin{enumerate}
\item[(1)] $|S_{N_0}|\le |I|$
\item[(2)] If $J$ is an interval and $n\in\bbN$, then the (Lebesgue) measure of 
the union of intervals $I_{n+1,j}$ which have nonzero intersection with $J$ is 
at most $|J|+2\bigl(\tfrac {2\pi}\beta  D_1^{-1}\ga^{-(n+1)}\bigr)$. 
\item[(3)] For any $j$ we have $\tfrac {2\pi} \beta D_1^{-1}\ga^{-(n+1)}\le |I_{n,j}| 
D_2(\ga D_1)^{-1}$. 
\end{enumerate} 
Here (2) and (3) follow from \eqref{A.2}. The 
net result is 
\[ 
|S_{N+N_0}|\le |I|\left(4a_i+6\frac{D_2}{\ga D_1}\right)^l\left(\frac 12-2a_i+2\frac{D_2}{\ga D_1}\right)^{N-l}. 
\]

Since $S_{N+N_0}$ is a union of intervals $I_{N+N_0+1,j}$, their number is (by 
\eqref{A.2}) at most 
\[
|I|\left(4a_i+\frac{6D_2}{\ga D_1}\right)^l
\left(\frac 12-2a_i+\frac{2D_2}{\ga D_1}\right)^{N-l}
\frac{\beta D_2}{2\pi}\ga^{N+N_0+1}. 
\]
Therefore $A_{N+N_0}$ can be covered by at most
\begin{align*}
\allowdisplaybreaks
\sum_{l=0}^{\lfloor(1-\eps)N\rfloor} c_0Ne^{-\frac{\eps^2}2 N} & \binom Nl 2^{N-l}
|I|\left(4a_i+\frac{6D_2}{\ga D_1}\right)^l
\left(\frac 12-2a_i+\frac{2D_2}{\ga D_1}\right)^{N-l}\frac{\beta D_2}{2\pi}\ga^{N+N_0+1} 
\\ & \le cN e^{-\frac{\eps^2}2 N}\ga^N \sum_{l=0}^N \binom Nl 
\left(4a_i+\frac{6D_2}{\ga D_1}\right)^l\left(\frac 12-2a_i+\frac{2D_2}{\ga D_1}\right)^{N-l}2^{N-l} 
\\ & =cN \left( \ga e^{-\frac{\eps^2}2}\left(1+\frac {10D_2}{\ga D_1}\right)\right)^N
\end{align*}
intervals $I_{N+N_0+1,j}$. Let us denote their union $A_{N+N_0}'$.

Now $\bigcup_{N=N_1}^\infty A_{N+N_0}'$ contains $A$, and by \eqref{A.2} it is a $\bigl( \tfrac{2\pi}\beta D_1^{-1}\ga^{-N_1-N_0-1} \bigr)$-cover. This yields
\begin{align*} 
\allowdisplaybreaks 
h^\alpha(A) &\le \lim_{N_1\to\infty}\sum_{N=N_1}^\infty 
cN \left( \ga e^{-\frac{\eps^2}2}\left(1+\frac {10D_2}{\ga D_1}\right)\right)^N \ga^{-(N+N_0+1)\alpha}
\\ &= \lim_{N_1\to\infty}\sum_{N=N_1}^\infty cN \left( \ga^{1-\alpha} e^{-\frac{\eps^2}2}\left(1+\frac {10D_2}{\ga D_1}\right)\right)^N.
\end{align*}
Hence if
\[
\ga^{1-\alpha} e^{-\frac{\eps^2}2}\left(1+\frac {10D_2}{\ga D_1}\right)<1,
\]
we get $h^\alpha(A)=0$. This is the case for any 
\[
\alpha>\alpha(\eps)\equiv 1-\frac{\eps^2-2\ln\left(1+\frac{10D_2}{\ga D_1}\right)}{2\ln(\ga)},
\]
and so $\dim(A)\le\alpha(\eps)$. If $\eps^2> 2\ln(1+\tfrac {10}\ga D_2D_1^{-1})$, then this is smaller than 1.

As mentioned earlier, it follows that $\dim(A_p^\eps)\le \alpha(\eps)$. Notice that
$\alpha(\eps)$ does not depend on $p$. Thus if
\[
A^\eps_X\equiv \left\{k\Big|\lims_{N\to\infty} \frac {\sum_{n=1}^N X_n(k)}N> \eps\right\},
\]
then obviously
\[
A_X^\eps\subseteq\bigcup_{p=1}^\infty A_{p}^\eps,
\]
and so $\dim(A^\eps_X)\le\alpha(\eps)$. Since the same result holds for 
\[
B^\eps_X\equiv \left\{k\Big|\limi_{N\to\infty} \frac {\sum_{n=1}^N X_n(k)}N< -\eps\right\},
\]
we can take $A^\eps\equiv A^\eps_X\cup B^\eps_X$. By the discussion at the beginning, this completes 
the proof.

We conclude with a generalization of {\it Theorem \ref{TA.1}}. We denote $x_\pm\equiv \max\{\pm x,0\}$.

\begin{theorem} \lb{TA.3}
{\it Theorem \ref{TA.1}\/} remains valid if we take
\[
F(k)\equiv 
\lims_{N\to\infty}\frac 1N \left|\sum_{n=1}^N G\bigl(f_n(k)\bigr) \right| 
\]
where $G:\bbR\to\bbR$ satisfies the following conditions: 
\begin{enumerate}
\item[(1)] $G$ is continuous and piecewise $C^1$;
\item[(2)] $G(x)=G(x+2\pi)$ and $\int_0^{2\pi} G(x)\,dx = 0$;
\item[(3)] There are $0=x_1<x_2< \dots <x_m=2\pi$ such that $G$ is monotone on 
$[x_j,x_{j+1}]$, and $G(x_{j+1})=0$ whenever $G(x_j)\neq 0$.
\end{enumerate} 
\end{theorem}

\noindent
{\it Remarks.}
1. The $x_j$'s are the zeros and local maxima and minima of $G$ in $[0,2\pi]$. Between two zeros $G$ has only one local extreme.
\smallskip

\noindent
2. In fact, we obtain
\begin{equation} \lb{A.5} 
\dim(A^\eps)
\le 1- \frac{ \left( \frac{\eps}{\|G_+\|_\infty+\|G_-\|_\infty}-\frac \pi 2 \|G'\|_\infty
\frac{\del_1+\del_2}{1-\del_1} \right)^2 
- 2\ln \left( 1+ \frac{M}\ga \frac{1+\del_2}{1-\del_1} \right) } {2\ln(\ga)} 
\end{equation}
whenever $\eps>\tfrac \pi 2 \|G'\|_\infty \bigl( \|G_+\|_\infty+\|G_-\|_\infty \bigr) (\del_1+\del_2)(1-\del_1)^{-1}$, 
with $M$ being the sum of twice the number of zeros and four times the larger of the numbers of local 
maxima and local minima of $G$. If in addition $G(\pi +x)=-G(\pi -x)$, then 
$\eps(\|G_+\|_\infty+\|G_-\|_\infty)^{-1}$ in \eqref{A.5} is replaced by just 
$\eps\|G\|_\infty^{-1}$.
\smallskip

\noindent
3. So, for instance, for $G(x)=\sin(x)$ we have $M=2\cdot 3+4\cdot 1=10$ as in {\it Theorem \ref{TA.1}}.
\smallskip

\begin{proof}[Proof outline.]
We first assume $G(\pi +x)=-G(\pi -x)$. In that case we proceed as before, but this time 
the functions $Y_{n,i}$ can take the value 0 on as many intervals as $G$ has zeros, 
and the value $1$ (resp. $-1$) on as many as $G$ has maxima (resp. minima). 
That is why $10$ is replaced by $M$.

If now $G$ is not odd with respect to $\pi$, we notice that \eqref{A.3} does not
guarantee $Y_n$ to have zero average. However, using the fact that $G$ has zero
average, and that any upper/lower Riemann sum is larger/smaller than the Riemann
integral, we can construct $Y_n$ with $\int_0^{2\pi} Y_n(k) \,dk=0$, 
$\|X_n-Y_n\|_\infty\le \tfrac 1p$ and such that $pY_n$ only takes integer 
values. The real price has to be paid when defining $Y_{n,i}$. If one wants them 
to have zero average, take only values 0 and $+1/-1$, and only on as many intervals as $G$ has 
zeros and local maxima/minima, then one may need to take $\approx 
p(\|G_+\|_\infty+\|G_-\|_\infty)$ of them. Since their sum is still just $pY_n$, 
this time we obtain 
\[ 
A_p^\eps\subseteq\bigcup_{i=1}^{p} A_p^{\eps_1}(i)
\]
with $\eps_1\equiv \eps(\|G_+\|_\infty+\|G_-\|_\infty)^{-1}$. This finishes the proof.
\end{proof}

\end{document}